\documentclass[12pt]{iopart}
\usepackage{iopams}  
\expandafter\let\csname equation*\endcsname\relax
\expandafter\let\csname endequation*\endcsname\relax
\usepackage{amsmath}
\usepackage{cite}
\usepackage{graphicx}
\usepackage{epsfig,color}
\usepackage{amssymb,eufrak}
\usepackage{amsthm}
\usepackage{enumerate}
\usepackage{hyperref}
\usepackage{braket}
\usepackage[dvipsnames]{xcolor}
\usepackage{tikz}
\usepackage{scrextend}

\definecolor{greencustom}{rgb}{0,0.75,0}

\begin{document}

\title{Digital quantum simulation of lattice gauge theories in three spatial dimensions}

\date{\today}

\author{Julian Bender$^1$, Erez Zohar$^1$, Alessandro Farace$^1$ and J. Ignacio Cirac$^1$}
\address{$^1$Max-Planck-Institut f\"ur Quantenoptik, Hans-Kopfermann-Stra\ss e 1, 85748 Garching, Germany.}

\begin{abstract}
In the present work, we propose a scheme for digital formulation of lattice gauge theories with dynamical fermions in 3+1 dimensions. All interactions are obtained as a stroboscopic sequence of two-body interactions with an auxiliary system. This enables quantum simulations of lattice gauge theories where the magnetic four-body interactions arising in two and more spatial dimensions are obtained without the use of perturbation theory, thus resulting in stronger interactions compared with analogue approaches. The simulation scheme is applicable to lattice gauge theories with either compact or finite gauge groups. The required bounds on the digitization errors in lattice gauge theories, due to the sequential nature of the stroboscopic time evolution, are provided. Furthermore, an implementation of a lattice gauge theory with a non-abelian gauge group, the dihedral group $D_{3}$, is proposed employing the aforementioned simulation scheme using ultracold atoms in optical lattices.
\end{abstract}

\section{Introduction}
Gauge theories lie at the core of fundamental physics; the standard model of particle physics - describing electromagnetic, weak and strong interactions - is based on the principle of gauge invariance \cite{eidelman2004review}. It requires introducing additional degrees of freedom, the gauge fields, to the matter fields: force carriers, mediating interactions between matter particles. If the coupling is small enough, perturbative expansions allow calculations up to arbitrary accuracy, as in QED (Quantum Electrodynamics). In some quantum field theories the coupling depends on the energy scale ($\textit{running coupling}$) \cite{peskin1995introduction, gross1973asymptotically}, and thus there are regimes where perturbation theory is not valid, e.g. QCD (Quantum Chromodynamics) at low energies. In such non-perturbative regimes only special methods can produce meaningful results. \\ 
The most common approach so far has been lattice gauge theory \cite{wilson1974confinement,kogut1979introduction}. The idea is to discretize space (or spacetime) to construct a framework in which numerical tools could be applied - with Monte Carlo methods being the most prominent ones \cite{smit2002introduction}. In spite of their success (e.g. calculation of the low-energy hadronic spectrum of QCD \cite{aoki2014review}), there are limitations which are inherent to Monte Carlo simulations of lattice gauge theories. A major one is the $\textit{sign problem}$, which prevents investigations in fermionic systems in finite chemical potential scenarios \cite{troyer2005computational}. As a consequence, corresponding phases in quantum field theories still remain relatively unexplored, e.g. the quark-gluon plasma or the color-superconducting phase of QCD \cite{mclerran1986physics,kogut2003phases}. Another drawback of these simulations is that they take place in a Euclidean spacetime, thus making real-time dynamics inaccessible and preventing, for example, the study of non-equilibrium phenomena. \\
One approach to overcome these obstacles is quantum simulation \cite{cirac2012goals,georgescu2014quantum}. The idea is to build a highly controllable quantum system serving as a platform for simulations of another quantum system. In particular, quantum simulations of lattice gauge theories \cite{zohar2015quantum,wiese2013ultracold} have been proposed using various quantum devices, such as ultracold atoms in optical lattices \cite
{jaksch1998cold,jaksch2005cold,bloch2008many}, trapped ions \cite{cirac1995quantum,leibfried2003quantum} or superconducting qubits \cite{you2003quantum,van1996one}. While the simulated models can be distinguished by features like the gauge group (abelian or non-abelian), the matter content (dynamical or static) or the dimension \cite{banerjee2013atomic, banerjee2012atomic,zohar2013simulating,zohar2011confinement,zohar2012simulating, gonzalez2017quantum,notarnicola2015discrete,zohar2013cold,stannigel2014constrained,zohar2017digital,zohar2016digital,hauke2013quantum,yang2016analog,tagliacozzo2013simulation,tagliacozzo2013optical,marcos2013superconducting,marcos2014two,mezzacapo2015non,zohar2013quantum,kasper2016schwinger,dutta2017toolbox,zohar2013topological,kosior2014simulation,wiese2014towards,walter2015dynamical}, there are also differences in the proposed simulation scheme. The first one - based on an idea of Feynman \cite{Feynman1982} - is to use a quantum computer (i.e. single and two qubit gates) to simulate the dynamics after Trotterization. The second one is an analogue approach: By appropriate engineering of the interactions, the Hamiltonian of the simulating system is exactly mapped to the desired one (which can be adiabatically changed), leading to an exact time evolution. The third one is a hybrid of both (e.g. \cite{jane2003simulation}), where the time evolution is Trotterized but the different terms of the Hamiltonian are implemented using an analogue simulation, instead of quantum gates.
It is important to note that the first simulation scheme will probably need quantum error correction, whereas the other two may not. Using the scheme suggested by Feynman, a trapped ion based quantum simulation of a lattice gauge theory was implemented in 2016 \cite{martinez2016real}, allowing the observation of real-time dynamics in the Schwinger model, (1+1) dimensional QED. However, the simulation involved only four ions and it remains a big challenge to scale up such a system as it involves the construction of a quantum computer. In this work, we will focus on the third option.\\ 
The main challenges of a quantum simulation of lattice gauge theories are threefold: First of all, to simulate dynamical matter, the simulating system must include fermionic degrees of freedom. Unlike in other quantum devices where fermionic statistics is imposed on spin degrees of freedom through Jordan-Wigner transformations, fermionic degrees of freedom occur naturally in ultracold atomic systems, as one can work directly with fermionic atomic species. This is beneficial in particular when dealing with two or more spatial dimensions. Second, gauge invariance, as the characteristic symmetry of lattice gauge theories, is not manifested naturally by the candidate quantum simulators. In analogue simulation schemes, where the degrees of freedom and the Hamiltonian of the investigated theory get exactly or approximately mapped onto the simulating system, local gauge invariance can be obtained either as a low-energy effective symmetry \cite{zohar2011confinement,zohar2012simulating,banerjee2012atomic} or by an exact mapping to an internal symmetry, like e.g. hyperfine angular momentum conservation \cite{zohar2013quantum,kasper2016schwinger}. Although the analogue approach works in one dimension (in particular as demonstrated by an ultracold atom experiment currently set up to study the Schwinger model \cite{kasper2017implementing}), it becomes problematic when considering the third requirement. The lattice gauge theory Hamiltonians in two or more spatial dimensions typically contain four-body interactions (the magnetic plaquette interactions). In the current analogue simulation schemes, this four-body term is realized only in fourth-order perturbation theory \cite{zohar2013quantum}, thus leading to weak interactions and posing a major challenge on the way to higher dimensional quantum simulations of lattice gauge theories. \\
This problem can be circumvented using the following concept: By introducing an auxiliary degree of freedom and entangling it with the physical degrees of freedom, the four-body interactions can be decomposed exactly as a sequence of simpler two-body interactions, resulting in stronger interactions compared to analogue simulation schemes. Because of the sequential nature of the entangling operations, during which all other interactions must be frozen, a stroboscopic time evolution is required. The time evolution is therefore decomposed into smaller pieces according to Trotter's formula: $e^{-itH}= \lim_{N \to \infty} ( \prod_{j} e^{-itH_{j}/N})^{N}$ \cite{trotter1959product}. This method has already been proposed in 2+1 dimensions to construct a digital scheme for lattice gauge theories with arbitrary gauge groups \cite{zohar2017digital}. A concrete quantum simulation with ultracold atoms has been proposed for the groups $\mathbb{Z}_{2}$ and  $\mathbb{Z}_{3}$ \cite{zohar2017digital,zohar2016digital}. \\ 
In this work we extend this proposal of an algorithm digitizing lattice gauge theories with arbitrary gauge groups to 3+1 dimensions. This is an important step towards the simulation of phenomena occurring in nature. To study the accuracy of the digital scheme, a thorough analysis of the digitization (Trotter) error is conducted. Another important goal is the simulation of gauge theories with non-abelian gauge groups. The second part of this work is therefore devoted to an ultracold atom based implementation of a lattice gauge theory with the simple non-abelian gauge group $D_{3}$, following the general algorithm presented in the first part. \\     
This paper is organized as follows: First, a brief lattice gauge theory background will be provided, with an emphasis on the Hamiltonian formulation used later on for quantum simulation. In the second section the digital algorithm enabling quantum simulation of lattice gauge theories with dynamical fermions in three dimensions will be described. Afterwards, improved bounds on the digitization errors in lattice gauge theories will be given. In the last section, possible implementations based on ultracold atoms will be discussed, in particular the implementation of a lattice gauge theory with the dihedral gauge group $D_{3}$, by exploiting its semidirect product structure. 

\section{Hamiltonian formulation of lattice gauge theories} \label{chapter2}

\begin{figure} 
\centering
\begin{tikzpicture}[
  line join=round,
  y={(-0.86cm,0.36cm)},x={(1cm,0.36cm)}, z={(0cm,1cm)},
  arr/.style={-latex,ultra thick,line cap=round,shorten <= 1.5pt}, line width=0.03cm
]
\def\Side{3}
\coordinate (A1) at (0,0,0);
\coordinate (A3) at (0,\Side,0);
\coordinate (A4) at (0,0,\Side);
\coordinate (A2) at (\Side,0,0);
\coordinate (B3) at (0,-\Side,0);
\coordinate (B4) at (0,0,-\Side);
\coordinate (B2) at (-\Side,0,0);
\draw (A3) --  node[right=0.25cm, above, black] {$\Theta_{g}^{L}(\mathbf{x},2)$} (A1);
\draw (A1) -- node[right=0.4cm, black]{$\Theta_{g}^{L}(\mathbf{x},1)$} (A2);
\draw (A1) -- node[right,black] {$\Theta_{g}^{L}(\mathbf{x},3)$} (A4);
\draw (A1) --  node[right=0.4cm, black] {${\Theta_{g}^R}^{\dagger}(\mathbf{x}-\mathbf{2},2)$} (B3);
\draw (A1) -- node[right=0.35cm, below, black]{${\Theta_{g}^R}^{\dagger}(\mathbf{x}-\mathbf{1},1)$} (B2);
\draw (A1) -- node[right,black] {${\Theta_{g}^R}^{\dagger}(\mathbf{x}-\mathbf{3},3)$} (B4);
\node[below] at (A1) {$\check{\theta}_{g}^{\dagger}(\mathbf{x})$};
\node[right,above] at (A2) {$\mathbf{x}+\mathbf{1}$};
\node[left,above] at (A3) {$\mathbf{x}+\mathbf{2}$};
\node[above] at (A4) {$\mathbf{x}+\mathbf{3}$};
\end{tikzpicture}
\caption{The local gauge transformation $\Theta_{g}(\mathbf{x})$, acting on the vertex $\mathbf{x}$ and adjacent links (shown here in three dimensions): $\check{\theta}_{g}^{\dagger}(\mathbf{x})$ acts on the fermionic Fock space at vertex $\mathbf{x}$, taking into account the staggered structure of the fermions. The three links $(\mathbf{x},k)$ emanating from vertex $\mathbf{x}$ are transformed by left transformations $\Theta_{g}^{L}$, whereas the incoming links $(\mathbf{x}-\mathbf{k},k)$ are transformed by right transformations ${\Theta_{g}^R}^{\dagger}$.}
\label{Gaugetrafo}
\end{figure}
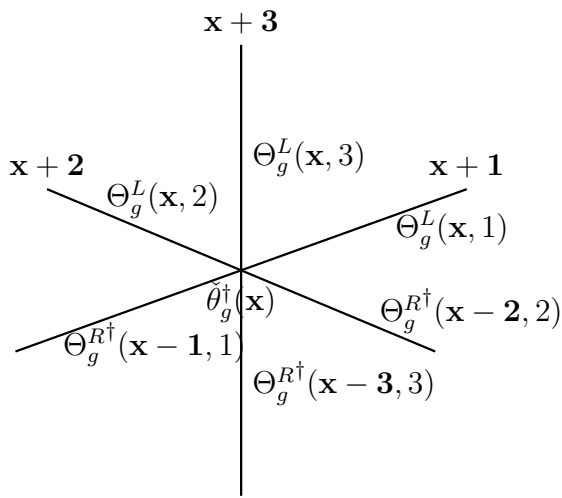

Lattice gauge theories can be formulated in a Hamiltonian framework exhibiting a continuous time coordinate, as first proposed by Kogut and Susskind \cite{kogut1975hamiltonian}. The lattice consists of $d$ spatial dimensions, where the matter fields are placed on the vertices $\mathbf{x} \in \mathbb{Z}^d$  and the gauge fields reside on the links $(\mathbf{x},k)$ (where $k \in \{1,..,d \}$ denotes the direction in which the link points). \\
Since the matter particles are allowed to tunnel and thus their number is not conserved locally, the states on the vertices are described by elements of a fermionic Fock space. Assuming the gauge group $G$ to be either compact or finite, we label its irreducible representations by $j$ and represent the matter fields by spinors $\psi_{m}^{\dagger \hspace{1pt} j}$, where $m$ denotes the components of $j$. Their behavior under group transformations, implemented by the unitary operator $\theta_{g}$, is (summing over repeated indices):
\begin{equation}
\begin{aligned}
&\theta_{g} \psi_{m}^{\dagger \hspace{1pt} j} \theta_{g}^{\dagger} = \psi_{n}^{\dagger \hspace{1pt} j}D_{nm}^{j}(g) \\
\end{aligned}
\end{equation}
where $D^{j}_{nm}(g)$ is the irreducible unitary representation $j$ of the group element $g$. We will work with $\textit{staggered fermions}$ \cite{susskind1977lattice}, distributing the Lorentz components of the spinor over neighboring lattice sites such that occupied even sites will correspond to particles and vacant odd sites to anti-particles. The Dirac spinor is then regained in a continuum limit. The gauge transformations $\check{\theta}_{g}$ of staggered fermions are related to $\theta_{g}$ by
\begin{equation}
\check{\theta}_{g} (\mathbf{x}) = \begin{cases} 
\theta_{g} & \text{for} \hspace{5pt} \mathbf{x} \in e \\
\theta_{g}  \det (D({g^{-1}})) & \text{for} \hspace{5pt} \mathbf{x} \in o 
\end{cases}
\end{equation}
with e (o) denoting the even (odd) sublattice. We can define a state $\ket{D}$ invariant under the above transformation (analogous to the Dirac sea in the continuum) where all odd sites are fully occupied and all even sites are vacant. \\
The other physical ingredients, the gauge degrees of freedom, are described by a tensor product of local Hilbert spaces on the links. The elements of each single link Hilbert space can be expressed in the \textit{group element states} $\{\ket{g}\}_{g \in G}$. The group G can act on it in two ways, corresponding to left ($L$) and right ($R$) transformations:
\begin{equation}
\begin{aligned}
\Theta_{g}^{L} \ket{h} =\ket{g^{-1}h}, \hspace{20pt} \Theta_{g}^{R} \ket{h}= \ket{hg^{-1}}
\end{aligned}
\end{equation}
We define the group element operator $U$, a matrix of operators acting on the link Hilbert space:

\begin{equation}
U_{mn}^{j} =\int \! D_{mn}^{j}(g) \ket{g} \bra{g} \, \mathrm{d}g
\end{equation}
where for continuous groups $\mathrm{d}g$ is understood as the group (Haar) measure, whereas for discrete groups the integral reduces to a sum over the group elements. \\
The hermitian conjugate of $U$ in the Hilbert space and in matrix space are related by  

\begin{equation} \label{hermi}
(U_{mn}^{j})^{\dagger} = \int dg \ket{g} \bra{g} \bar{D}_{mn}^{j}(g) = \int dg \ket{g} \bra{g} D_{nm}^{j  \dagger}(g)=(U^{j  \dagger})_{nm} 
\end{equation}
The group element operators obey the following rules under group transformations: 

\begin{equation}
\begin{aligned}
\Theta_{g}^{L} \hspace{1pt} U_{mn}^{j} {\Theta_{g}^{L}}^{\dagger}= D_{mm'}^{j}(g) \hspace{1pt} U_{m'n}^{j}, \hspace{30pt}
\Theta_{g}^{R} \hspace{1pt} U_{mn}^{j} {\Theta_{g}^{R}}^{\dagger}= U_{mn'}^{j} \hspace{1pt} D_{n'n}^{j}(g)
\end{aligned}
\end{equation}
(the $j$ will be omitted in the following as only one fixed representation $j$ is considered; generalization to more representations is straightforward). With these definitions at hand we can define a local gauge transformation which acts on all degrees of freedom intersecting at a vertex. It depends on a group element $g$ which itself can depend on the position (see Fig. \ref{Gaugetrafo} for illustration):

\begin{equation}
\Theta_{g}(\mathbf{x})=\prod_{k=1..d} \left( \Theta_{g}^{L}(\mathbf{x},k) {\Theta_{g}^R}^{\dagger}(\mathbf{x}-\mathbf{k},k) \right) \check{\theta}_{g}^{\dagger}(\mathbf{x})
\end{equation}
where $\mathbf{k}$ is the unit vector in $k$-direction. A state $\ket{\psi}$ is therefore said to be gauge-invariant if 

\begin{equation}
\Theta_{g} (\mathbf{x}) \ket{\psi} = \ket{\psi} , \hspace{20pt} \forall \mathbf{x},g
\end{equation}

Introducing the dual basis to the group element states, the \textit{representation basis} $\{ \ket{jmn} \}$, connected by the relation $\braket{g|jmn}=\sqrt[]{\frac{dim(j)}{|G|}} D_{mn} (g)$ (with $j$ labeling irreducible representations and $m$,$n$ the components under left and right transformations), we can define a gauge-invariant "empty" state for the whole lattice, including matter and gauge fields:

\begin{equation} \label{startstate}
\ket{0}\equiv \ket{D} \bigotimes_{\mathrm{links}} \ket{000}
\end{equation}
where $\ket{000}$ is a singlet state of the gauge fields in the representation basis, corresponding to the trivial representation. All other gauge invariant states can be obtained by acting with gauge invariant operators on this trivial state. A conventional lattice gauge theory Hamiltonian consists of four such types of terms: 

\begin{enumerate}
\item {The $\textit{magnetic Hamiltonian}$ \\
One can obtain gauge invariant operators by taking products of $U$-operators along closed paths. The shortest such possible path is a plaquette, characterized by two directions $k$ and $l$ ($k<l$ and $l \in \{2,..,d \}$). Adding over all pairs of $k$ and $l$ for every vertex $\mathbf{x}$, one may construct: 
\begin{equation}
H_{B}= \lambda_{B} \sum_{\mathbf{x},k < l} \mathrm{Tr}  \left( U(\mathbf{x},k) U(\mathbf{x}+\mathbf{k},l) U^{\dagger}(\mathbf{x}+\mathbf{l},k) U^{\dagger}(\mathbf{x},l)  \right) + H.c.
\end{equation}
This term is called $\textit{magnetic Hamiltonian}$ as it corresponds to the magnetic energy in the continuum limit of the Yang-Mills cases.}

\item{The $\textit{electric Hamiltonian}$ \\
\begin{equation}
\begin{aligned}
H_{E}&=\lambda_{E} \sum_{\mathbf{x},k} h_{E}(\mathbf{x},k) \\
\mathrm{with} \hspace{10pt} h_{E}(\mathbf{x},k)&= \sum_{j,m,n} f(j) \ket{jmn} \bra{jmn}
\end{aligned}
\end{equation}
The correspondence with the electric field becomes clear for the case of $G=U(1)$ where - if we set $f(j)=j^2$ - the Hamiltonian is just a sum over the square of the electric field of all links. Similarly, for $SU(2)$ $f(j)=j(j+1)$ corresponding to $\mathbf{J}^2$.}

The two terms above involve only gauge fields. They both add up to 
\begin{equation}
H_{KS}=H_{B}+H_{E} \hspace{2pt}, 
\end{equation}
a generalized version of the $\textit{Kogut-Susskind Hamiltonian}$ for lattice gauge theories with compact gauge groups \cite{zohar2015formulation}. 

\item{The $\textit{fermionic mass Hamiltonian}$ \\
Introducing staggered fermions gives rise to the following staggered mass term: 

\begin{equation}
H_{M}= M \sum_{\mathbf{x}} (-1)^{\mathbf{x}} \psi^{\dagger}_{n} (\mathbf{x}) \psi_{n} (\mathbf{x}) 
\end{equation}
where the alternating sign comes from the Dirac sea picture: particles on even sites and anti-particles on odd sites.
}

\item{The $\textit{gauge-matter Hamiltonian}$ \\
The last term is a fermionic hopping term minimally coupled to the gauge fields in a gauge invariant way:  
\begin{equation}
H_{GM}=\lambda_{GM} \sum_{\mathbf{x},k} \psi^{\dagger}_{m} (\mathbf{x}) U_{mn}(\mathbf{x},k) \psi_{n}(\mathbf{x}+\mathbf{k}) +H.c.
\end{equation}
}
\end{enumerate}
The total Hamiltonian we want to simulate in the following chapters is the sum of all four pieces. The state defined in (\ref{startstate}) is the non-interacting vacuum: the ground state of $H_{E}+H_{M}$.

\section{Digital algorithm for the quantum simulation of lattice gauge theories in three dimensions} \label{algo}

Interactions in typical quantum simulation platforms are usually two-body, e.g. atomic collisions in ultracold atomic setups or spin-spin interactions in trapped ion setups. Three-and four body processes are strongly suppressed on the relevant experimental timescales, making it much harder to map the Hamiltonian of the simulated model onto the simulating system, if the former includes interactions of more than two bodies. This is particularly relevant for lattice gauge theories since magnetic interactions are four-body terms (see Sec. \ref{chapter2}). For the purpose of quantum simulation of lattice gauge theories it is therefore desirable to design a scheme in which interactions involving three and more constituents can be rewritten as exact sequences of only two-body interactions. In this way, the energy scale associated to plaquette interactions is not limited by perturbative arguments (as in previous proposals) and the simulation can give access to a bigger region of the phase diagram.\\
One approach to this problem is based on the idea of using an auxiliary degree of freedom that gets entangled with the physical degrees of freedom and mediates their interactions. In the following, we will briefly present an isometry which formalizes this idea (it is sometimes referred to as \textit{stator} \cite{reznik2002remote, zohar2017half}). We anticipate that in this new framework the time evolution has to be realized stroboscopically due to the sequential nature of the entangling operations with the auxiliary system. Therefore, a digital algorithm based on Trotter's formula will be designed to simulate lattice gauge theories in three spatial dimensions, using only two-body interactions. This corresponds to the hybrid simulation scheme discussed in the introduction, where the time evolution is Trotterized but the individual parts of the Hamiltonian are still implemented by an analogue simulation. In the last section, bounds on the Trotter error will be provided.  

\subsection{Isometries} \label{statorsection}

We consider two Hilbert spaces: $\mathcal{H}_{A}$ representing the "physical" degrees of freedom, where the interaction is supposed to be implemented, and $\mathcal{H}_{B}$ representing the auxiliary degrees of freedom (sometimes called control in the following). We denote the operators acting on the Hilbert space $\mathcal{H}$ by $\mathcal{O}(\mathcal{H})$. An isometry $S$ can then be defined, mapping $\mathcal{H}_{A} \to \mathcal{H}_{A} \otimes \mathcal{H}_{B}$, which can be created by a unitary $\mathcal{U}_{AB} \in \mathcal{O}(\mathcal{H}_{A} \otimes \mathcal{H}_{B})$ acting on some initial state $\ket{\text{in}_{B}} \in \mathcal{H}_{B}$: 
\begin{equation}
S=\mathcal{U}_{AB}\ket{\text{in}_{B}} \in \mathcal{O}(\mathcal{H}_{A}) \otimes \mathcal{H}_{B}
\end{equation}
This can be viewed as an entangling operation between the physical and the auxiliary degrees of freedom. If this entangling procedure is chosen in a certain way, operations on the physical Hilbert space can be implemented by acting only on the auxiliary state. Assume we want to realize a Hamiltonian $H \in \mathcal{O}(\mathcal{H}_{A})$ in the physical Hilbert space. For that, we need to create an isometry S and a hermitian operator $H' \in \mathcal{H}_{B}$ in the auxiliary Hilbert space in such a way that the following relation holds: 

\begin{equation}
\begin{aligned}
H'S=SH
\end{aligned}
\end{equation}
An analogue relation for the time evolution follows directly, since $\hspace{3pt} {H'}^nS=SH^n$: 

\begin{equation}
\begin{aligned}
e^{-iH't}S=Se^{-iHt}
\end{aligned}
\end{equation}
Therefore, by creating such an isometry and acting with $H'$ on the control, we obtain the desired time evolution of the physical state $\ket{\psi_{A}}$: 

\begin{equation}
\begin{aligned}
e^{-iH't} \mathcal{U}_{AB} \ket{\psi_{A}} \ket{\text{in}_{B}}=e^{-iH't} S \ket{\psi_{A}} = S e^{-iHt} \ket{\psi_{A}}
\end{aligned}
\end{equation}
The evolved physical state is still entangled with the auxiliary state which means that one can either perform another operation using the isometry $S$ or disentangle both states. This would lead to a product state with the auxiliary state going back to its initial state: 

\begin{equation}
\begin{aligned}
\mathcal{U}_{AB}^{\dagger} e^{-iH't} \mathcal{U}_{AB} \left( \ket{\psi_{A}} \otimes \ket{\text{in}_{B}} \right)= \ket{\text{in}_{B}} \otimes e^{-iHt} \ket{\psi_{A}}
\end{aligned}
\end{equation}

\subsection{The three-dimensional algorithm}

\begin{figure}
\centering
\includegraphics[width=0.5 \textwidth]{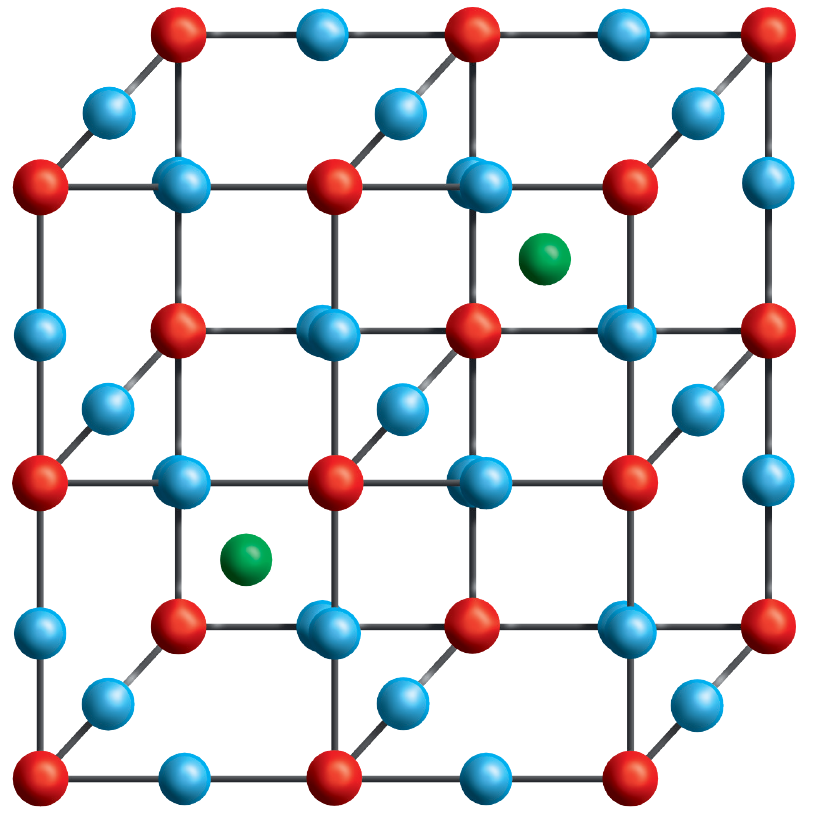}
\caption{The physical system consists of the gauge fields residing on the links (blue) and the matter fields on the vertices (red). The auxiliary degrees of freedom (green) are located in the center of every second cube (either even or odd).}
\label{3DImp}
\end{figure}

In this section we discuss an algorithm to simulate the lattice gauge theory Hamiltonian in three spatial dimensions. We start from the lattice model described in Sec. \ref{chapter2}. To create plaquette and gauge-matter interactions by means of isometries, we introduce an auxiliary degree of freedom in the middle of every second cube (either all even or odd ones) and assign to it a Hilbert space $\widetilde{\mathcal{H}}$ isomorphic to the Hilbert spaces on the links (see Fig \ref{3DImp}). Then, the lattice gauge theory Hamiltonian is split up into several parts which are implemented independently and sequentially: 

\begin{equation}
H_{LGT}=H_{E}+H_{M}+ \sum_{i=1}^{6} H_{B,i}+ \sum_{j=1}^6 H_{GM,j} \label{HamTrot}
\end{equation}
where we explicitly distinguish gauge-matter interactions taking place along different directions and in odd or even cubes, as well as plaquette interactions corresponding to the different plaquettes of a unit cube (therefore we get a sum of six terms in both cases). The desired time evolution $e^{-i t H_{LGT}}$ is then approximated by a Trotterized time evolution consisting of $N$ steps: $e^{-i t H_{LGT}} \sim ( \prod_{j} e^{-itH_{j}/N})^{N}$, where $H_j$ is any of the terms appearing in ~\eqref{HamTrot}. While electric and mass terms can be treated easily using only the physical degrees of freedom, the plaquette and gauge-matter terms are further decomposed as a suitably chosen sequence of simpler interactions mediated by the auxiliary systems. This sequence will then be executed in parallel for all cubes where auxiliary degrees of freedom are located. However, since for the gauge-matter interactions the individual parts of this sequence do not commute for adjacent links, we have to place the auxiliary d.o.f. in every second cube to avoid undesired interactions. The exact decompositions will be given in the next sections. \\

\subsubsection{Plaquette interactions} \label{algoplaq}
Since we put auxiliary atoms in every second cube, we can not realize all plaquette interactions at once and we split them up in the following way: 

\begin{equation}
\begin{aligned}
H_{B} &= \sum_{\mathbf{x}} \left( \lambda_{B} \mathrm{Tr}(U(\mathbf{x},1)U(\mathbf{x}+\mathbf{1},2)U^{\dagger}(\mathbf{x}+\mathbf{2},1)U^{\dagger}(\mathbf{x},2)) + H.c. \right) \\
& \hspace{17pt}  + \left(\lambda_{B} \mathrm{Tr}(U(\mathbf{x},3)U(\mathbf{x}+\mathbf{3},1)U^{\dagger}(\mathbf{x}+\mathbf{1},3)U^{\dagger}(\mathbf{x},1)) +H.c. \right)\\
& \hspace{17pt}  +\left( \lambda_{B} \mathrm{Tr}(U(\mathbf{x},2)U(\mathbf{x}+\mathbf{2},3)U^{\dagger}(\mathbf{x}+\mathbf{3},2)U^{\dagger}(\mathbf{x},3)) +H.c. \right)  \\
&\equiv \sum_{\mathbf{x}} \left( H_{B,1}(\mathbf{x})+H_{B,2}(\mathbf{x})+H_{B,3}(\mathbf{x}) \right)\\
&=\sum_{\mathbf{x} \hspace{2pt}\mathrm{even}} \left(H_{B,1}(\mathbf{x})+H_{B,2}(\mathbf{x})+H_{B,3}(\mathbf{x}) \right) +\sum_{\mathbf{x} \hspace{2pt}\mathrm{odd}} \left( H_{B,1}(\mathbf{x})+H_{B,2}(\mathbf{x})+H_{B,3}(\mathbf{x}) \right)\\
&\equiv H_{B,1e}+H_{B,2e}+H_{B,3e}+H_{B,1o}+H_{B,2o}+H_{B,3o} \\
\end{aligned}
\end{equation}
It is important to mention that the six magnetic terms commute, therefore $e^{-i \tau H_B} = \prod_{j} e^{-i \tau H_{B,je}} e^{-i \tau H_{B,jo}}$ and this splitting does not affect the error of the Trotter approximation \eqref{HamTrot}. To implement each term we will use the isometry  

\begin{equation} \label{groupstator}
S_{i}=\int \! \mathrm{d}g  \ket{g}_{i} \bra{g}_{i} \otimes \ket{\tilde{g}} 
\end{equation}
where the first Hilbert space belongs to the gauge field, residing on link $i$, and the second one to the aforementioned auxiliary degree of freedom in the center of the cube. It fulfills the relation 

\begin{equation} \label{eigenoperatorgroup}
\widetilde{U} S_{i}=S_{i} U_{\mathrm{link} \hspace{1pt} i}
\end{equation}
allowing to realize operations on the link $i$ through the auxiliary degree of freedom. The isometry $S_{i}$ can be created by the unitary 

\begin{equation}  \label{groupstatorcreate}
\mathcal{U}_{i}=\int \! \mathrm{d}g  \ket{g}_{i} \bra{g}_{i} \otimes {\Theta_{g}^L}^{\dagger} 
\end{equation}
acting on the initial state $\ket{\widetilde{\text{in}}}=\ket{\tilde{e}}$. We repeat similar entangling operations $\mathcal{U}_{i}$ (or $\mathcal{U}_{i}^{\dagger}$) for the three other links of the plaquette (e.g. the links 1,2,3,4 of cube $\mathbf{x}$, see Fig. \ref{Defi}) and obtain a plaquette isometry of the form

\begin{equation}
S_{\square}^{1234} (\mathbf{x}) =\mathcal{U}^{1234}_{\square} (\mathbf{x}) \ket{\widetilde{\text{in}}}=\mathcal{U}_{1} (\mathbf{x})\mathcal{U}_{2}(\mathbf{x}) \mathcal{U}_{3}^{\dagger} (\mathbf{x}) \mathcal{U}_{4}^{\dagger} (\mathbf{x}) \ket{\widetilde{\text{in}}}
\end{equation}
The crucial part is that it fulfills the relation
\begin{equation} \label{eigenoperator}
\mathrm{Tr}(\widetilde{U}(\mathbf{x})+\widetilde{U}^{\dagger}(\mathbf{x})) \hspace{2pt} S_{\square}^{1234}(\mathbf{x})=S_{\square}^{1234}(\mathbf{x}) \hspace{2pt} \mathrm{Tr}(U_{1}(\mathbf{x})U_{2}(\mathbf{x})U_{3}^{\dagger}(\mathbf{x})U_{4}^{\dagger}(\mathbf{x})+ H.c.)
\end{equation}
i.e. acting locally with 
\begin{equation}
\widetilde{H}_{B}(\mathbf{x})=\lambda_{B} \mathrm{Tr}(\widetilde{U}(\mathbf{x})+\widetilde{U}^{\dagger}(\mathbf{x}))
\end{equation}
on the control of cube $\mathbf{x}$ enables us to realize the magnetic time evolution for this plaquette. The required sequence acting on the plaquette state $\ket{\psi_{1234}}$, the tensor product of the four link states, and the auxiliary state $\ket{\widetilde{\text{in}}}$ is 
\begin{equation} \label{sequenceHB}
\mathcal{U}_{\square}^{1234 \hspace{1pt}\dagger} (\mathbf{x}) \hspace{2pt}  e^{-i\widetilde{H}_{B}(\mathbf{x}) \tau} \hspace{2pt} \mathcal{U}_{\square}^{1234} (\mathbf{x}) \ket{\psi_{1234}} \ket{\widetilde{\text{in}}}=\ket{\widetilde{\text{in}}} e^{-i H_{B,1}(\mathbf{x})\tau}   \ket{\psi_{1234}}
\end{equation}
The other two plaquette terms associated to cube $\mathbf{x}$ can be created in the same manner but with different isometries. Using the abbreviations for the gauge field operators defined according to Fig. \ref{Defi}, we need to replace the isometry $S_{\square}^{1234} (\mathbf{x})$ by $S_{\square}^{5671} (\mathbf{x}) = \mathcal{U}_{\square}^{5671} (\mathbf{x}) \ket{\widetilde{\text{in}}}$ (green, dashed plaquette), or $S_{\square}^{5894} (\mathbf{x}) = \mathcal{U}_{\square}^{5894} (\mathbf{x}) \ket{\widetilde{\text{in}}}$ (blue, dotted plaquette). Applying the sequence from (\ref{sequenceHB}) gives then rise to the time evolution of the physical state under $H_{B,2}(\mathbf{x})$, or $H_{B,3}(\mathbf{x})$.

\begin{figure} [h]
\begin{minipage}{0.7 \textwidth}
\begin{tikzpicture}[
  line join=round,
  y={(-0.86cm,0.36cm)},x={(1cm,0.36cm)}, z={(0cm,1cm)},
  arr/.style={-latex,ultra thick,line cap=round,shorten <= 1.5pt}, line width=0.03cm
]
\def\Side{3.5}
\coordinate (A1) at (0,0,0);
\coordinate (A2) at (0,\Side,0);
\coordinate (A3) at (\Side,\Side,0);
\coordinate (A4) at (\Side,0,0);
\coordinate (B1) at (0,0,\Side);
\coordinate (B2) at (0,\Side,\Side);
\coordinate (B3) at (\Side,\Side,\Side);
\coordinate (B4) at (\Side,0,\Side);

\draw (A2) -- (B2) -- (B1);
\draw (B1) -- (B4) -- (A4);
\draw[red,line width=0.07cm] (A3) -- node[above, black]{$U_{3}$}(A2) -- node[below, black]{$U_{4}$} (A1);
\draw[blue, dotted,line width=0.07cm] (A2) -- (A1);
\draw[red,line width=0.07cm] (A1) -- node[below, black]{$U_{1}$} (A4);
\draw[greencustom, dash pattern=on 6pt off 3pt,line width=0.07cm] (A1) -- (A4);
\draw[blue, dotted, line width=0.07cm] (B2) -- node[above, black]{$U_{8}$}(B1);
\draw[greencustom, dash pattern=on 6pt off 3pt, line width=0.07cm] (B1) -- node[above, black]{$U_{6}$}(B4);
\draw (B4) -- (B3) -- (B2);
\draw[blue, line width=0.07cm] (A1) -- node[right] {} (B1);
\draw[greencustom, dash pattern=on 6pt off 3pt, line width=0.07cm] (A1) -- node[right, black] {$U_{5}$} (B1);
\draw[blue, dotted, line width=0.07cm] (A2) -- node[left, black]{$U_{9}$}(B2);
\draw[greencustom, dash pattern=on 6pt off 3pt, line width=0.07cm] (A4) -- node[right, black]{$U_{7}$}(B4);
\draw (A3) -- (B3);
\draw[red,line width=0.07cm] (A3) -- node[above, black]{$U_{2}$}(A4);



\node[below] at (A1) {$\mathbf{x}$};
\node[below] at (A2) {$\mathbf{x}+\mathbf{2}$};
\node[below] at (A3) {};
\node[below] at (A4) {$\mathbf{x}+\mathbf{1} $};
\node[above] at (B1) {$\mathbf{x}+\mathbf{3}$};
\node[above] at (B2) {};
\node[above] at (B3) {};
\node[above] at (B4) {};

\end{tikzpicture}
\end{minipage}
\begin{minipage}[r]{0.2\textwidth}
\begin{equation*}
\begin{aligned}
U_{1}(\mathbf{x})&\equiv U(\mathbf{x},1) \\
U_{2}(\mathbf{x})&\equiv U(\mathbf{x}+\mathbf{1},2) \\
U_{3}(\mathbf{x})&\equiv U(\mathbf{x}+\mathbf{2},1) \\
U_{4}(\mathbf{x})&\equiv U(\mathbf{x},2) \\
U_{5}(\mathbf{x})&\equiv U(\mathbf{x},3) \\
U_{6}(\mathbf{x})&\equiv U(\mathbf{x}+\mathbf{3},1) \\
U_{7}(\mathbf{x})&\equiv U(\mathbf{x}+\mathbf{1},3)  \\
U_{8}(\mathbf{x})&\equiv U(\mathbf{x}+\mathbf{3},2)  \\
U_{9}(\mathbf{x})&\equiv U(\mathbf{x}+\mathbf{2},3)  \\
\end{aligned}
\end{equation*}
\end{minipage}

\caption{There are three different plaquette terms associated to every vertex $\mathbf{x}$: $H_{B,1}(\mathbf{x})$ (red, solid plaquette), $H_{B,2}(\mathbf{x})$ (green, dashed plaquette) and $H_{B,3}(\mathbf{x})$ (blue, dotted plaquette). Each term involves four gauge field operators $U$, abbreviated as above for a convenient description.}
\label{Defi}
\end{figure} 
We can now formulate an algorithm to implement the whole plaquette interactions. We start with the controls placed in the center of every even cube and do the following three steps:
\begin{enumerate}
\item{Create the isometry: Let all the controls interact with all the gauge fields on links of type 4 and create the unitary $\prod\limits_{\mathbf{x} \hspace{2pt} \mathrm{even}} \mathcal{U}_{4}^{\dagger} (\mathbf{x})$. Repeat similar processes with links 3, 2 and 1 to obtain the unitaries $\prod\limits_{\mathbf{x} \hspace{2pt} \mathrm{even}} \mathcal{U}_{3}^{\dagger} (\mathbf{x})$, $\prod\limits_{\mathbf{x} \hspace{2pt} \mathrm{even}} \mathcal{U}_{2} (\mathbf{x})$, $\prod\limits_{\mathbf{x} \hspace{2pt} \mathrm{even}} \mathcal{U}_{1} (\mathbf{x})$. In total, we get: $\prod\limits_{\mathbf{x} \hspace{2pt} \mathrm{even}} \mathcal{U}_{1} (\mathbf{x})\mathcal{U}_{2}(\mathbf{x}) \mathcal{U}_{3}^{\dagger} (\mathbf{x}) \mathcal{U}_{4}^{\dagger} (\mathbf{x})=\prod\limits_{\mathbf{x} \hspace{2pt} \mathrm{even}} \mathcal{U}_{\square}^{1234} (\mathbf{x})$.}
\item{Act on the controls with the Hamiltonian $\sum\limits_{\mathbf{x} \hspace{2pt} \mathrm{even}} \widetilde{H}_{B}(\mathbf{x})$ for time $\tau$, resulting in the time evolution $\prod\limits_{\mathbf{x} \hspace{2pt} \mathrm{even}} e^{-i\widetilde{H}_{B}(\mathbf{x}) \tau}$.}
\item{In the last step, undo the isometry by creating the inverse of the first step, i.e. $\prod\limits_{\mathbf{x} \hspace{2pt} \mathrm{even}} \mathcal{U}_{\square}^{1234 \hspace{1pt}\dagger} (\mathbf{x}) $.}
\end{enumerate}
The above procedure is applied to a state $\ket{\psi}\ket{\widetilde{\text{in}}}$. Thanks to  relation (\ref{sequenceHB}) we obtain:
\begin{equation} \label{timeevmag}
\prod\limits_{\mathbf{x} \hspace{2pt} \mathrm{even}} \mathcal{U}_{\square}^{1234 \hspace{1pt}\dagger} (\mathbf{x}) e^{-i\widetilde{H}_{B}(\mathbf{x}) \tau} \mathcal{U}_{\square}^{1234} (\mathbf{x}) \ket{\psi}\ket{\widetilde{\text{in}}}= \ket{\widetilde{\text{in}}} e^{-iH_{B,1e}\tau}  \ket{\psi} \equiv \ket{\widetilde{\text{in}}} W_{B,1e}\ket{\psi}
\end{equation}
We repeat the procedure with the two isometries $S_{\square}^{5671}$ and $S_{\square}^{5894}$. In this way we create $W_{B,2e}=e^{-iH_{B,2e}\tau}$ and $W_{B,3e}=e^{-iH_{B,3e}\tau}$. The same steps are then repeated with the auxiliary degrees of freedom moved to the center of the odd cubes so that we can implement $W_{B,1o},W_{B,2o},W_{B,3o}$. Since all pieces of the magnetic Hamiltonian commute, this sequence gives us exactly the magnetic time evolution: $W_{B,1e}W_{B,2e}W_{B,3e}W_{B,1o}W_{B,2o}W_{B,3o}=W_{B}=e^{-i\tau H_{B} }$. \\

\subsubsection{Gauge-Matter interactions} \label{algogauge}
\begin{figure} 
\centering

\begin{tikzpicture}[
  line join=round,
  y={(-0.86cm,0.36cm)},x={(1cm,0.36cm)}, z={(0cm,1cm)},
  arr/.style={-latex,ultra thick,line cap=round,shorten <= 1.5pt}, line width=0.03cm
]
\def\Side{3.5}
\coordinate (A1) at (0,0,0);
\coordinate (A2) at (0,\Side,0);
\coordinate (A3) at (\Side,\Side,0);
\coordinate (A4) at (\Side,0,0);
\coordinate (B1) at (0,0,\Side);
\coordinate (B2) at (0,\Side,\Side);
\coordinate (B3) at (\Side,\Side,\Side);
\coordinate (B4) at (\Side,0,\Side);

\draw (A1) -- (A2);
\draw (A1) -- (B1);
\draw (A3) -- (A2);
\draw[greencustom, dash pattern=on 6pt off 3pt, line width=0.07cm] (A2) --  node[left=0.4cm, below, black] {$U_{ml}(\mathbf{x},2)$} (A1);
\draw[red, line width=0.07cm] (A1) -- node[right=0.4cm, below, black]{$U_{mn}(\mathbf{x},1)$} (A4);
\draw (B2) -- (B1) -- (B4) -- (B3) -- cycle;
\draw[blue, dotted, line width=0.07cm] (A1) -- node[right,black] {$U_{mk}(\mathbf{x},3)$} (B1);
\draw (A2) -- (B2);
\draw (A4) -- (B4);

\draw (A3) -- (B3);
\draw (A3) -- (A4);



\node[below] at (A1) {$\psi_{m}^{\dagger}(\mathbf{x}) $};
\node[left] at (A2) {$\psi_{l}^{\dagger}(\mathbf{x}+\mathbf{2}) $};
\node[below] at (A3) {};
\node[right] at (A4) {$\psi_{n}^{\dagger}(\mathbf{x}+\mathbf{1}) $};
\node[above=0.1cm] at (B1) {$\psi_{k}^{\dagger}(\mathbf{x}+\mathbf{3}) $};
\node[above] at (B2) {};
\node[above] at (B3) {};
\node[above] at (B4) {};

\end{tikzpicture}
\caption{There are three gauge-matter terms associated to every vertex $\mathbf{x}$, corresponding to the three links emanating from this vertex: $H_{GM}(\mathbf{x},1)$ (red, solid link), $H_{GM}(\mathbf{x},2)$ (green, dashed link) and $H_{GM}(\mathbf{x},3)$  (blue, dotted link). Each interaction consists of two fermions $\psi_{m}^{\dagger}(\mathbf{x})$ and $\psi_{n}^{\dagger}(\mathbf{x}+\mathbf{k})$ located at the endpoints of the links and the gauge field operator $U_{mn}(\mathbf{x},k)$ on the link.}
\label{Link}
\end{figure}
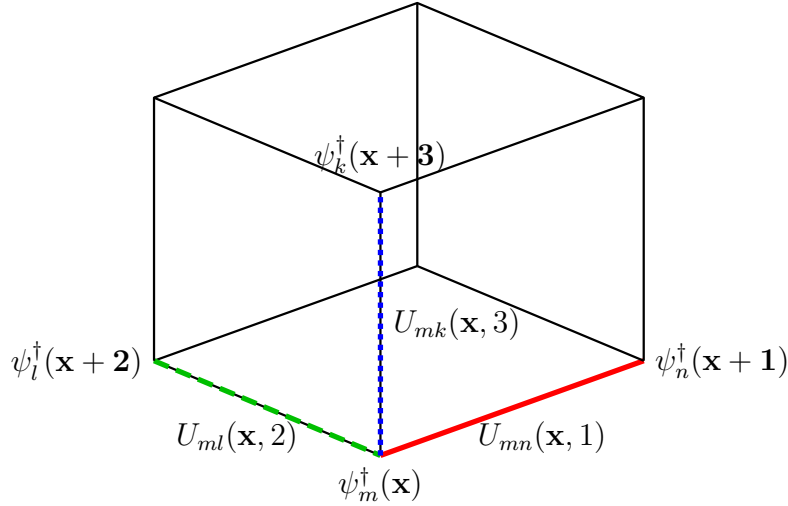

After expressing the four-body plaquette interactions as a sequence of two-body interactions, we want to obtain the gauge-matter interactions in a similar way. We need again to split up the relevant Hamiltonian terms into parts suitable for implementation:

\begin{equation}
\begin{aligned}
H_{GM}&= \sum_{\mathbf{x}} \sum_{k=1}^{3} \lambda_{GM} \psi^{\dagger}_{m}(\mathbf{x})U_{mn}(\mathbf{x},k)\psi_{n}(\mathbf{x}+\mathbf{k})+ H.c. \\
&= \sum_{\mathbf{x} \hspace{2pt}\mathrm{even}} \left( H_{GM}(\mathbf{x},1)+H_{GM}(\mathbf{x},2)+H_{GM}(\mathbf{x},3) \right) \\ 
&\hspace{4pt}+\sum_{\mathbf{x} \hspace{2pt}\mathrm{odd}} \left( H_{GM}(\mathbf{x},1)+H_{GM}(\mathbf{x},2)+H_{GM}(\mathbf{x},3) \right)\\
&\equiv H_{GM,1e}+H_{GM,2e}+H_{GM,3e}+H_{GM,1o}+H_{GM,2o}+H_{GM,3o}
\end{aligned}
\end{equation}
An important ingredient for rewriting these interactions as two-body terms is the following unitary operation, entangling the fermion at vertex $\mathbf{x}$ and the gauge field on link $(\mathbf{x}, k)$:  

\begin{equation}
\mathcal{U}_{W}(\mathbf{x},k)= e^{i Z_{mn}(\mathbf{x},k) \psi_{m}^{\dagger} (\mathbf{x})  \psi_{n} (\mathbf{x})  }
\end{equation}
where $Z_{mn}=-i(\log_{\mathrm{mat}}(U))_{mn}$, and the logarithm is taken only in matrix space (well-defined since the matrix elements commute). Its meaning becomes more apparent if we assume the gauge group $G$ to be compact; then, we obtain   

\begin{equation}
\mathcal{U}_{W}(\mathbf{x},k)=e^{i\hat{\phi}^a(\mathbf{x},k)\psi_{m}^{\dagger}(\mathbf{x})T_{mn}^{a}\psi_{n}(\mathbf{x})}=e^{i  \hat{\phi}^a Q^{a} }
\end{equation}
an interaction of the "vector potential" operator $\hat{\phi}^{a}$ with the fermionic charge $Q^{a}$. It can therefore be interpreted as a fermionic transformation whose parameter is an operator acting on the gauge field. The idea is now to use this transformation to map a pure fermionic tunneling term into the desired gauge-matter interactions, as
\begin{equation}
\mathcal{U}_{W}(\mathbf{x},k) \psi^{\dagger}_{n} (\mathbf{x}) \mathcal{U}_{W}^{\dagger}(\mathbf{x},k) = \psi_{m}^{\dagger} (\mathbf{x}) U_{mn} (\mathbf{x},k) 
\end{equation}
Thus, defining the fermionic tunneling Hamiltonian as 

\begin{equation}
H_{t}(\mathbf{x},k)=\lambda_{GM} (\psi^{\dagger}_{m}(\mathbf{x})\psi_{m}(\mathbf{x}+\mathbf{k})+H.c.)  
\end{equation}
allows writing the Hamiltonian $H_{GM}$ as: 

\begin{equation} \label{GMInt}
H_{GM}(\mathbf{x},k)= \mathcal{U}_{W}(\mathbf{x},k) H_{t}(\mathbf{x},k) \mathcal{U}_{W}^{\dagger}(\mathbf{x},k) 
\end{equation}
Since every fermion is connected to six links in three dimensions we have to split up the process in six steps as described in the beginning. We start by realizing $H_{GM,1e}$, i.e. $H_{GM}(\mathbf{x},1)$ for all even links (see Fig. $\ref{Link}$). We apply the following sequence:
\begin{enumerate}
\item{Let the gauge degrees of freedom interact with the fermions at the beginning of the link to obtain the unitary: $\prod\limits_{\mathbf{x} \hspace{2pt} \mathrm{even}} \mathcal{U}_{W}^{\dagger}(\mathbf{x},1)$ .}
\item{Allow tunneling on these links for time $\tau$: $\prod\limits_{\mathbf{x} \hspace{2pt} \mathrm{even}} e^{-iH_{t}(\mathbf{x},1)\tau}$ .}
\item{Let the link degrees interact again with the fermions to generate: $\prod\limits_{\mathbf{x} \hspace{2pt} \mathrm{even}} \mathcal{U}_{W}(\mathbf{x},1)$ .}
\end{enumerate}
This gives us in total
\begin{equation}
\prod\limits_{\mathbf{x} \hspace{2pt} \mathrm{even}} \mathcal{U}_{W}(\mathbf{x},1) e^{-iH_{t}(\mathbf{x},1)\tau} \mathcal{U}_{W}^{\dagger}(\mathbf{x},1) =e^{-i\sum\limits_{\mathbf{x} \hspace{2pt} \mathrm{even}} H_{GM}(\mathbf{x},1)\tau} \equiv W_{GM,1e}
\end{equation}
By applying a similar sequence for the other links of the cube, we can create $W_{GM,2e},W_{GM,3e},W_{GM,1o},W_{GM,2o},W_{GM,3o}$. \\

Using isometries, there is an alternative way of realizing the gauge-matter interactions. It requires more steps but on the other hand does not require interactions between the physical degrees of freedom as all of them are mediated by the auxiliary degrees of freedom. The sequence goes as follows: 
\begin{enumerate}
\item{Let the controls - initially placed in all even cubes in the state $\ket{\widetilde{\text{in}}}=\ket{\tilde{e}}$ - interact with the gauge links $U_{1}$ according to (\ref{groupstatorcreate}) to create the isometry $S_{1}$: $\prod\limits_{\mathbf{x} \hspace{2pt} \mathrm{even}} \mathcal{U}_{1} (\mathbf{x})$ .}
\item{Let the control interact with the fermion at vertex $\mathbf{x}$ to realize the interaction $\hspace{4pt} \prod\limits_{\mathbf{x} \hspace{2pt} \mathrm{even}} \widetilde{\mathcal{U}}^{\dagger}_{W} (\mathbf{x},1)$ which is the same interaction as  $\mathcal{U}_{W}^{\dagger} (\mathbf{x},1)$ but between the control and the fermion $\psi_{m}(\mathbf{x})$. Due to the properties of the isometry $S_{1}$ the interaction between the control and the fermion will translate into an interaction between the fermion and the link.}
\item{Afterwards, allow for pure tunneling between the fermions which gives rise to $\prod\limits_{\mathbf{x} \hspace{2pt} \mathrm{even}} e^{-iH_{t}(\mathbf{x},1)\tau}$ .}
\item{Following (\ref{GMInt}), apply $\widetilde{\mathcal{U}}_{W} (\mathbf{x},1)$ for all even cubes which is again realized by an interaction between the control and the fermion $\psi_{m}(\mathbf{x})$: $\prod\limits_{\mathbf{x} \hspace{2pt} \mathrm{even}} \widetilde{\mathcal{U}}_{W} (\mathbf{x},1)$ .}
\item{Finally, we have to undo the isometry between the control and the gauge field: $\prod\limits_{\mathbf{x} \hspace{2pt} \mathrm{even}} \mathcal{U}_{1}^{\dagger} (\mathbf{x})$ .}
\end{enumerate}
The resulting sequence - applied to some physical state $\ket{\psi}$ and the auxiliary state $\ket{\widetilde{\text{in}}}$ - is: 
\begin{equation}
\begin{aligned}
\prod\limits_{\mathbf{x} \hspace{2pt} \mathrm{even}} \mathcal{U}_{1}^{\dagger} (\mathbf{x}) \widetilde{\mathcal{U}}_{W} (\mathbf{x},1) e^{-iH_{t}(\mathbf{x},1)\tau} \widetilde{\mathcal{U}}^{\dagger}_{W} (\mathbf{x},1) \mathcal{U}_{1} (\mathbf{x}) \ket{\psi}\ket{\widetilde{\text{in}}}=\ket{\widetilde{\text{in}}}e^{-i H_{GM,1e}}\ket{\psi}=\ket{\widetilde{\text{in}}}W_{GM,1e}\ket{\psi}
\end{aligned}
\end{equation}
We repeat a similar procedure for all other links in the cube which gives us $W_{GM,2e}, W_{GM,3e}, W_{GM,1o}, W_{GM,2o}, W_{GM,3o}$.

\subsubsection{Other parts of the Hamiltonian} \label{other}
The electric part $W_{E}=e^{-iH_{E}\tau}$ and the matter part $W_{M}=e^{-iH_{M}\tau}$ are local terms of our Hamiltonian and thus one can implement them by acting locally on the physical degrees of freedom. \\
We can now write down the whole sequence for a time step $\tau$ (combining commuting magnetic terms to $W_{B}$):
\begin{equation} \label{sequence}
\begin{aligned}
W_{\tau}=W_{M}W_{E}W_{GM,3o}W_{GM,2o}W_{GM,1o}W_{GM,3e}W_{GM,2e}W_{GM,1e}W_{B} \\
\end{aligned}
\end{equation}
It's important to notice that all time evolutions in the above sequence are individually gauge-invariant. Therefore, errors coming from the digitization do not break gauge-invariance.

\subsection{Error bounds for Trotterized time evolutions in lattice gauge theory} \label{errorbounds}
Although the approximated Trotter evolution has the correct gauge symmetry, it is still important to analyze how much it deviates from the desired exact time evolution. In this section we derive bounds for the Trotter error, according to the digitization scheme presented in the previous section. We focus on the standard Trotter formula (the first order formula) and the second order formula which gives a better approximation without major changes in the implementation. We do not consider higher order formulas, because they would require more experimental effort in the sense that the tunability of the experimental parameters would have to be much more flexible and the number of operations required for a single time step would increase exponentially with the order of the approximation \cite{berry2007efficient}. The first order formula \cite{trotter1959product} is of the form 

\begin{equation}
\mathcal{U}_{N}(t)=(\prod_{j} e^{-iH_{j}\frac{t}{N}})^N
\end{equation}
For which, using the operator norm, the difference to the physical time evolution $\mathcal{U}(t)=e^{-itH}$ can be bounded by \cite{kalos2012monte,binder1992monte,rebbi1983lattice}:

\begin{equation} \label{Firstorder}
\begin{aligned}
\left\Vert \mathcal{U}(t)-\mathcal{U}_{N}(t) \right\Vert \le \frac{t^2}{2N} \sum_{j<k} \left\Vert \left[H_{j},H_{k}\right] \right\Vert
\end{aligned}
\end{equation}
To get a better scaling with the number of time steps we can apply the Trotterization sequence in reverse order after the usual Trotterized time evolution (second order formula) \cite{suzuki1991general}:

\begin{equation} \label{ZweiteOrdnung}
\mathcal{U}_{2,N}(t)= \left( e^{-iH_{1} \frac{t}{2N}} ... e^{-iH_{p-1} \frac{t}{2N}}e^{-iH_{p} \frac{t}{N}} e^{-iH_{p-1} \frac{t}{2N}} ... e^{-iH_{1} \frac{t}{2N}} \right)^N
\end{equation}
From an implementation point of view this decomposition can be realized straightforwardly once we know how to obtain the sequence for the first order. Following the proof in \cite{suzuki1985decomposition} adapted to unitary operators, an upper bound for the trotter error can be derived:

\begin{equation} \label{Secondorder}
\begin{aligned}
&\left\Vert\mathcal{U}(t)-\mathcal{U}_{2,N}(t) \right\Vert \\ 
=& \left\Vert e^{-itH}  - (e^{-iH_{1} \frac{t}{2N}} ... e^{-iH_{p-1} \frac{t}{2N}}e^{-iH_{p} \frac{t}{N}} e^{-iH_{p-1} \frac{t}{2N}} ... e^{-iH_{1} \frac{t}{2N}})^N  \right\Vert \\
\le&\frac{t^3}{12 N^2}  \sum_{k=1}^{p-1} \left\Vert [[H_{k},H_{k+1}+..+H_{p}],H_{k+1}+..+H_{p}] \right\Vert + \frac{1}{2} \left\Vert [[H_{k},H_{k+1}+..+H_{p}],H_{k}] \right\Vert
\end{aligned}
\end{equation}
Compared to the first order formula, the second order formula has an error which decreases faster with the number of time steps $N$ at the cost of a longer sequence. The experimental difficulty, however, is the same for both decompositions. \\
We can now specify these bounds for lattice gauge theories. This is an important task since an implementation of this digital scheme will have to balance experimental errors, which can break gauge-invariance and increase with the number of steps in the sequence, and errors caused by the digitization, which have the opposite behavior. Therefore, a precise bound of the Trotter error helps in finding the optimal number of steps, so that experimental errors do not accumulate unnecessarily and the chance of breaking gauge invariance is reduced as much as possible. \\
Since the different parts of the Hamiltonian can not be implemented simultaneously, they are split up in the digitized simulation scheme. Hence, for the computation of the trotter error we divide the Hamiltonian into these individual pieces, according to the Trotterized time evolution given in (\ref{sequence}). Generalizing to $d$ dimensions: 
\begin{equation} \label{Trott}
H_{LGT}=H_{B}+H_{E}+H_{M}+ \sum_{i=1}^{2d} H_{GM,i}
\end{equation}

\subsubsection{First order formula}
By inspection of ($\ref{Firstorder}$) we see that for an upper bound on the digitization error of the standard trotter formula, the commutators among all different parts of the Hamiltonian in ($\ref{Trott}$) have to be evaluated, as well as their norms. Since the derivations are very lengthy we will refer the interested reader to the Appendix. We provide here the final result: 

\begin{equation} \label{firstordertrott}
\begin{aligned}
&\|\mathcal{U}(t)-\mathcal{U}_{N}(t) \| \\ 
\le& \frac{t^2}{2N} \left( \| \left[H_{E}, H_{B} \right] \| + \| [H_{GM},H_{E}] \| + \| H_{GM},H_{M} \| + \sum_{k=1}^{2d-1} \sum_{j=k+1}^{2d} \| \left[H_{GM,j},H_{GM,k} \right] \|        \right) \\ 
=&\frac{t^2d_{U} \mathcal{N}_{\mathrm{links}}}{N} \left(  \lambda_{B} \lambda_{E} 4(d-1)\max_{j} |f(j)| + \lambda_{GM}  \lambda_{E} \max_{j} |f(j)| + M \lambda_{GM} + \lambda_{GM}^2 \frac{2d-1}{4} \right) 
\end{aligned}
\end{equation}
where $d$ is the number of spatial dimensions, $d_{U}$ the dimension of the representation of the group element operator $U$ and $\mathcal{N}_{\mathrm{links}}$ the number of links in the lattice. One might think that operator norms involving $H_{E}$ are unbounded but, since we either work with finite groups (whose number of irreducible representations is finite) or appropriate truncations of infinite gauge field Hilbert spaces, the expression $\max_{j} |f(j)|$ is finite, so that we always obtain sensible error bounds.

\subsubsection{Second order formula}
To bound the error of the second order formula we need to calculate nested commutators according to (\ref{Secondorder}). Details on their calculation can be found in the Appendix. We provide here the final result: 

\begin{equation} \label{secondordertrott}
\begin{aligned}
&\|\mathcal{U}(t)-\mathcal{U}_{2,N}(t) \|\\
\le& \frac{t^3  \mathcal{N}_{\mathrm{links}} d_{U}}{6 N^2} \left[ 16 \lambda_{E}\lambda_{B} \max_{j} |f(j)| (d-1) \left( 2\lambda_{E} \max_{j} |f(j)| +\lambda_{B} d_{U} (d-1)   \right) \right.\\
&\left.+ \lambda_{GM}\lambda_{E} \max_{j} |f(j)| \left( 2\lambda_{GM} d_{U} (2(2d-1)+1) + \lambda_{E} \max_{j} |f(j)|   \right) \right.\\
&\left.+ \lambda_{GM}M \left( 4d\lambda_{GM}+M \right) + \lambda_{GM}^3(2d-1) \left(\frac{1}{3} (4d-1)+\frac{1}{2} \right) \right]
\end{aligned}
\end{equation}
If we assume a cubic lattice with $L$ lattice sites per side we can express the number of links as: $\mathcal{N}_{\mathrm{links}}= d (L-1) L^{d-1}$. The upper bound shows that $N$ should scale as $N \sim L^{d/2}t^3$ which is somewhat bad since it considers a very general setting. If we restrict ourselves to the observation of intensive quantities we expect this scaling to be much better. However, there are observables in lattice gauge theories, e.g. Wilson loops, which do not fulfill this requirement and thus need to be bounded by more general estimates like the ones given above.

\section{Implementation of digital lattice gauge theories with ultracold atoms}
With this general scheme for the digital construction of three-dimensional lattice gauge theories at hand, we can turn to the implementation of some concrete examples with ultracold atoms. Typical gauge groups of interest are compact (e.g. $U(1)$), for which the link Hilbert spaces are infinite. A truncation of this Hilbert space is therefore required to make the quantum simulation feasible. Previous proposals have performed this truncation in the $\textit{representation basis}$ \cite{zohar2015quantum,wiese2013ultracold}. This procedure, however, spoils unitarity of the group element operators $U$ and prevents the use of isometries (see \ref{statorsection}). Thus, the Hilbert space of the gauge field should be truncated using $\textit{group element states}$ instead. A truncation of $U(1)$ in this sense is given by the finite groups $\mathbb{Z}_{N}$ which converge to $U(1)$ in the $N \to \infty$ limit. The digital quantum simulation of $\mathbb{Z}_N$ gauge theories has been studied in \cite{zohar2016digital,zohar2017digital}. We summarize below their main features, and then we build on these to tackle the simulation of simple non-abelian gauge models with dihedral symmetry given by the group $D_N$.  

\subsection{Implementation of lattice gauge theories with gauge group $\mathbb{Z}_{N}$}                                                                                                                       
Lattice gauge theories with a finite abelian gauge group play an important role as they approximate compact quantum electrodynamics. Since the Hilbert space of the gauge field is reduced to dimension $N$ if the gauge group $\mathbb{Z_{N}}$ is considered, ultracold atoms can be used to represent these gauge degrees of freedom. These $N$ states are labeled by $\ket{m}$ and we define unitary operators $P$ and $Q$ on them: 

\begin{equation}
\begin{aligned}
P^N&=Q^N=1 \\
PQP^{\dagger}&=e^{i\frac{2\pi}{N}} Q \\
Q \ket{m}&=\ket{m+1} \hspace{10pt}(\mathrm{cyclically})\\
P \ket{m}&=e^{i \frac{2 \pi}{N}m} \ket{m}
\end{aligned}
\end{equation}
Since the group is abelian, its representations are one dimensional and we need to consider a single fermionic species, $\psi^{\dagger}$, on the vertices. We can now define the Hamiltonian of $\mathbb{Z}_{N}$ lattice gauge theory with fermionic matter:

\begin{equation} \label{electricZN}
\begin{aligned}
&H_{E}=\lambda_{E} \sum_{\mathbf{x},k} \left( 1 - P(\mathbf{x},k) - P^{\dagger}(\mathbf{x},k) \right)\\
&H_{B}= \lambda_{B} \sum_{\mathbf{x},k<l} Q(\mathbf{x},k)Q(\mathbf{x}+\mathbf{k},l)Q^{\dagger}(\mathbf{x}+\mathbf{l},k)Q^{\dagger}(\mathbf{x},l) +H.c.  \\
&H_{M}= M \sum_{\mathbf{x}} (-1)^{\mathbf{x}} \psi^{\dagger} (\mathbf{x}) \psi (\mathbf{x}) \\
&H_{GM}= \lambda_{GM} \sum_{\mathbf{x},k} \psi^{\dagger} (\mathbf{x}) Q(\mathbf{x},k) \psi(\mathbf{x}+\mathbf{k}) +H.c. 
\end{aligned}
\end{equation}
Possible implementations for $\mathbb{Z}_{2}$  \cite{zohar2016digital} and $\mathbb{Z}_{3}$ \cite{zohar2017digital} with isometries have been discussed in two space dimensions. These proposals can be readily generalized to three dimensions following the scheme presented in the previous section. The matter content is represented by a fermionic atomic species whereas the gauge fields can be represented by a second atomic species with the appropriate ground state manifold, e.g. $F=1/2$ for $\mathbb{Z}_{2}$ or $F=1$ for $\mathbb{Z}_{3}$. Furthermore, auxiliary atoms must be trapped in the center of each second cube. These species are confined to the desired lattice geometry by suitable optical lattices and their interactions are realized by ultracold atomic scattering. Since the type of interactions appearing in two and three dimensions are the same, the implementation in three dimension follows closely the steps explained in \cite{zohar2016digital,zohar2017digital} and the reader should refer to the original references for more details. \\
Here we just report the bounds on the Trotter error that can be computed following the discussion in Sec. \ref{errorbounds}. In three dimensions and for the gauge group $\mathbb{Z}_{N}$, we obtain the first order formula (see (\ref{firstordertrott})) : 

\begin{equation}
\begin{aligned}
\|\mathcal{U}(t)-\mathcal{U}_{N}(t) \| \le\frac{3 t^2 (L-1)L^2}{N} \left( 16 \lambda_{B} \lambda_{E} + 2 \lambda_{GM}  \lambda_{E}  + M \lambda_{GM} + \lambda_{GM}^2 \frac{5}{4} \right) 
\end{aligned}
\end{equation}

and the second order formula (see (\ref{secondordertrott})): 
\begin{equation}
\begin{aligned}
\|\mathcal{U}(t)-\mathcal{U}_{2,N}(t) \| \le& \frac{t^3  (L-1) L^{2}}{ N^2} \left( 64 \lambda_{E}\lambda_{B}  ( 2\lambda_{E}  +\lambda_{B}    ) + 2\lambda_{GM}\lambda_{E} ( 11\lambda_{GM}  + \lambda_{E}    ) \right.\\
&+ \left. \lambda_{GM}M ( 6 \lambda_{GM}+\frac{1}{2} M ) +  \frac{125}{12} \lambda_{GM}^3     \right)
\end{aligned}
\end{equation}
Note that these formulas give a more accurate bound with respect to the original analysis in \cite{zohar2016digital,zohar2017digital}.

\subsection{Implementation of lattice gauge theories with a dihedral gauge group}
We now turn our attention to the implementation of simple non-abelian lattice gauge theories, with symmetry given by the dihedral group $D_{N}$ (with $N$ odd and $N \geq 3$ which converges in the large-N limit to $O(2)$). This symmetry group can be characterized by a set of rotations $R$ in a two-dimensional plane and reflections $S$ along a certain axis: 

\begin{equation}
D_{N} = \{ g= (p , m) \equiv R\left(2\pi/N \right)^p S^m | p \in [0,N-1 ) \hspace{2pt} \mathrm{and} \hspace{2pt} m \in \{ 0,1 \}  \} 
\end{equation}
The above notation already suggests that $D_{N}$ can be decomposed into a semidirect product of the abelian groups $\mathbb{Z}_{N}$ and $\mathbb{Z}_{2}$ corresponding to rotations and reflections: $D_{N} \simeq \mathbb{Z}_{N} \rtimes \mathbb{Z}_{2}$. It is thus useful to write the states of the gauge field Hilbert space as states living in the tensor product of an $N$-dimensional Hilbert space and a two-dimensional one, $\ket{p,m}= \ket{p} \otimes \ket{m} \in \mathcal{H}_{N} \otimes \mathcal{H}_{2}$. In the implementation, such a product Hilbert space can be realized by using two atoms with the appropriate hyperfine structure. If we choose to work with the smallest faithful irreducible representation of the group, we need two different fermionic components for the matter, denoted by $\psi_{1}$ and $\psi_{2}$, due to the non-abelian nature. Accordingly, the gauge field operators $U$ on the links become matrices of operators $U=e^{i \frac{2 \pi}{N} \hat{p} \sigma_{z}} \sigma_{x}^{\hat{m}}$ ($\hat{p}$ acts on $\mathcal{H}_{N}$ and $\hat{m}$ on $\mathcal{H}_{2}$; $\sigma_{x}$ and $\sigma_{z}$ act in matrix space). This allows us to write down the Hamiltonians 
\begin{equation}
\begin {aligned}
H_{B}&= \lambda_{B} \sum_{\mathbf{x},k<l} \mathrm{Tr}  \left( U(\mathbf{x},k) U(\mathbf{x}+\mathbf{k},l) U^{\dagger}(\mathbf{x}+\mathbf{l},k) U^{\dagger}(\mathbf{x},l)  \right) + H.c. \\
H_{GM}&=\lambda_{GM} \sum_{\mathbf{x},k} \begin{pmatrix}
\psi^{\dagger}_{1} (\mathbf{x}), &\psi^{\dagger}_{2}  (\mathbf{x})
\end{pmatrix} 
e^{i \frac{2 \pi}{N} \hat{p} \sigma_{z}} \sigma_{x}^{\hat{m}} (\mathbf{x},k) \begin{pmatrix}
\psi_{1}(\mathbf{x}) \\
\psi_{2}(\mathbf{x}) 
\end{pmatrix}   +H.c. \\
H_{M}&= M \sum_{\mathbf{x}} (-1)^{\mathbf{x}} \psi^{\dagger} (\mathbf{x}) \psi (\mathbf{x})= M \sum_{\mathbf{x}} (-1)^{\mathbf{x}} \left(\psi^{\dagger}_{1} (\mathbf{x}) \psi_{1} (\mathbf{x})+\psi^{\dagger}_{2} (\mathbf{x}) \psi_{2} (\mathbf{x}) \right) 
\end{aligned}
\end{equation} 
The last part, the electric Hamiltonian, takes its simplest form if the states in $\mathcal{H}_{2}$ are expressed in the usual group element states $\{ \ket{m} \} $ but the states of $\mathcal{H}_{N}$ in $\{ \ket{l}, l=0,..,N-1 \}$, the conjugate basis to $\{ \ket{p} \}$ (defined by ${ \braket{l|p}=\frac{1}{\sqrt[]{N}} e^{-i\frac{2 \pi}{N} lp} }$, see Appendix for details): 

\begin{equation}
\begin{aligned}
H_{E}&= \lambda_{E} \sum_{\mathbf{x},k} h_{E} (\mathbf{x},k) \\
\mathrm{with} \hspace{10pt} h_{E}(\mathbf{x},k)&=\frac{1}{2} \sum_{m,m'}   f_{r} \ket{0,m} \bra{0,m'} + \sum_{l= -(N-1)/2}^{(N-1)/2} \sum_{m}  f_{l} \ket{l,m} \bra{l,m} 
\label{ElDN}
\end{aligned}
\end{equation}
where $f_{r}$ and $f_{l}$ satisfy the condition $f_{l}=f_{-l} \hspace{2pt} \forall l$. $D_{N}$ lattice gauge theories do not have a meaningful large-$N$ limit (like $\mathbb{Z}_{N}$ with compact QED) as $O(2)$ is not a "conventional" lattice gauge theory and does not have a continuum limit. Thus, in principle the coefficients in \eqref{ElDN} can be chosen arbitrarily. However, it is convenient to identify the second term of \eqref{ElDN} with the electric energy of a $\mathbb{Z}_{N}$ lattice gauge theory (see above), and fix the coefficients accordingly.

\subsubsection{Simulating system}
\begin{figure} 
\centering
\includegraphics[width=0.8 \textwidth]{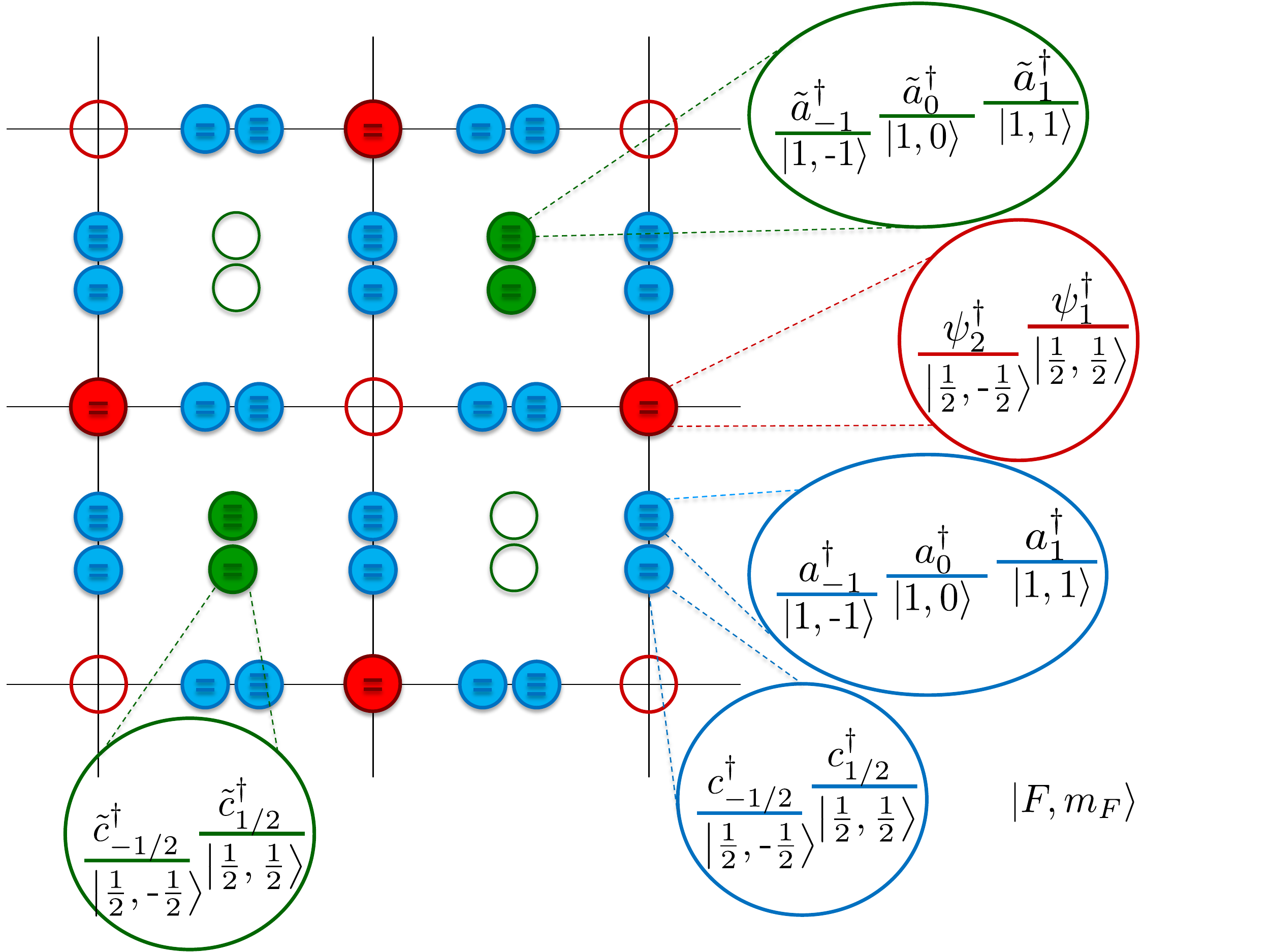}
\caption{The simulating system consists of one atomic species on the vertices representing the matter (red) and two for both the gauge fields (blue) on the links and the controls (green) located at the center of every second cube (projected into two dimensions for better visualization). The simulated degrees of freedom are encoded in the hyperfine structure of the atoms, i.e. either an $F=1$ or an $F=1/2$ multiplet. The alternating occupation of vertices with fermionic atoms shall illustrate the staggered fermion picture, in which this configuration corresponds to the non-interacting vacuum (see \textit{Dirac sea} in the continuum). The empty green circles indicate the need to move the auxiliary atoms between even and odd cubes.}
\label{d3}
\end{figure}

Our implementation scheme is in principle applicable to all dihedral groups but we focus here on the simplest case $D_{3}$ (isomorphic to the group of permutations $S_{3}$). We first discuss the system we will use as a platform to perform the quantum simulation. \\
For the simulation of the matter fields it is crucial to use fermionic atoms to obtain the correct commutation relations. A natural, minimal choice for the two fermionic d.o.f. $\psi_{1}$ and $\psi_{2}$ is to use the two internal levels of an atom with a $F=1/2$ hyperfine ground state. For example $\psi_{1}$ and $\psi_{2}$ can be associated with the $F=1/2$ multiplet in the following way:

\begin{equation}
\begin{aligned}
\psi_{1}^{\dagger}& \to \ket{F=1/2;m_{F}=1/2} \\
\psi_{2}^{\dagger}& \to \ket{F=1/2;m_{F}=-1/2} 
\end{aligned}
\end{equation}
These atoms must be trapped by a superlattice that allows to modulate the depth of the minima (to account for the staggering) and the tunneling rate between nearest neighbors (to switch tunneling on and off in the different steps of the Trotter sequence). \\
To simulate the gauge field and auxiliary Hilbert spaces, we will exploit the product structure as mentioned above: $\mathcal{H}_{aux} \simeq \mathcal{H}_{link} \simeq \mathcal{H}_{3} \otimes \mathcal{H}_{2}$. One convenient choice is to use two atomic species: a bosonic one with an $F_{3}=1$ hyperfine multiplet (the index 3 will label the three-level system) and a fermionic one with an $F_{2}=1/2$ multiplet (the index 2 will label the two-level system). In total, we need four different atomic species: two atoms trapped at the middle of each link, and two extra atoms (that must be addressed independently of the previous two) in the middle of each second cube. For the links, we identify: 

\begin{equation}
\begin{aligned}
\ket{p=0}&\equiv\ket{F_{3}=1,m_{F}=0} \hspace{50pt} \ket{m=0}\equiv\ket{F_{2}=1/2,m_{F}=1/2}\\
\ket{p=1}&\equiv\ket{F_{3}=1,m_{F}=1} \hspace{50pt} \ket{m=1}\equiv\ket{F_{2}=1/2,m_{F}=-1/2}\\
\ket{p=2}&\equiv\ket{F_{3}=1,m_{F}=-1} 
\end{aligned}
\end{equation}
Every state of the Hilbert space on the link can be obtained as a tensor product of the two multiplets, e.g. $\ket{p=1,m=1}=\ket{F_{3}=1,m_{F}=1} \otimes \ket{F_{2}=\frac{1}{2},m_{F}=-\frac{1}{2}}$. The corresponding creation operators on some link $(\mathbf{x},k)$ are described by $a^{\dagger}_{m_{F}}(\mathbf{x},k)$ with $m_{F}=-1,0,1$ for the three-level system and $c^{\dagger}_{m_{F}}(\mathbf{x},k)$ with $m_{F}=-1/2,1/2$ for the two-level system. It is useful to introduce unitary operators $P_{3},Q_{3}$ and $P_{2},Q_{2}$ acting respectively on the three-level and two-level atoms. They are defined as:

\begin{equation}
\begin{aligned}
P_{3} \ket{p} &= e^{i\frac{2\pi}{3}p} \ket{p}   \hspace{90pt}    P_{2} \ket{m} = (-1)^m  \ket{m}\\
Q_{3} \ket{p} &= \ket{p+1} (\text{cyclically}) \hspace{38pt}     Q_{2} \ket{m} = \ket{m+1} (\text{cyclically})\\
\end{aligned}
\end{equation}
The operators $P_3$, $Q_3$ fulfill the $\mathbb{Z}_{3}$ algebra whereas the operators $P_2$, $Q_2$ fulfill the $\mathbb{Z}_{2}$ algebra.\\
The Hilbert space of the auxiliary atoms has the same structure, and we label its states/operators with a tilde to distinguish them from the corresponding link quantities, i.e. we have states $\ket{\tilde{p}}$ and $\ket{\tilde{m}}$ and operators $\tilde{a}^{\dagger}_{m_{F}}(\mathbf{x})$ (with $m_{F}=-1,0,1$) and $\tilde{c}^{\dagger}_{m_{F}}(\mathbf{x})$ (with $m_{F}=-1/2,1/2$). \\ 
The link and auxiliary atoms must be trapped in the desired positions by arranging suitable optical potentials. The individual minima must contain exactly one atom and must be deep and well separated so that the dynamics is frozen (no tunneling, no interactions between nearest neighbors). When requested, the lattices must undergo a rigid translation so that specific pairs of atoms can overlap and interact via two-body scattering. The resulting setup - for convenience projected to two dimensions - is depicted in Fig. $\ref{d3}$. \\
All interactions between the constituents of the simulating system from above are in the form of two-body scattering. As will become clear in the following, we need to impose specific constraints on the scattering. First we want interactions that are diagonal in $m_{F}$ and do not change the internal level of the atoms. This can be achieved by lifting the degeneracy of the hyperfine multiplets such that transitions changing $m_{F}$ will cost energy. A possible way to do this is by introducing a uniform magnetic field which adds the following correction to the Hamiltonian (Zeeman shift): 

\begin{equation}
\begin{aligned}
H_{Z}= \mu_{B} g_{F} m_{F} B 
\end{aligned}
\end{equation}
where $\mu_{B}$ is the Bohr magneton and $g_{F}$ the hyperfine Lande factor. The energy splitting has to be different for different atomic species to avoid resonant exchanges, therefore we need to choose species with different Lande factors. Another possible approach to realize the different energy splittings is to address each species individually, for example exploiting the AC Stark effect. Second, at some point we need to modulate the interaction strengths depending on the internal level of the atoms. This can be achieved for example by spatially separating the different $m_{F}$ levels via a magnetic field gradient. The different $m_{F}$ levels will experience forces pointing in different directions and reach different equilibrium positions within the same potential well. By properly choosing the Lande factors of the atomic species and tuning the magnetic field gradient one can then tailor the overlap of the atomic Wannier wave functions (and hence their interaction strength) in an $m_{F}$-dependent way. \\
Below we discuss several details of the implementation scheme.

\subsubsection{Initial configuration and background Hamiltonian}

Before starting the simulation we should define the initial configuration of our simulating system. It is reached if all optical potentials are sufficiently deep and separated such that no tunneling  occurs and all atomic wave functions do not overlap. All minima of the auxiliary lattice are loaded with one atom in the group element state corresponding to the identity group element, i.e $\ket{\widetilde{\text{in}}}=\ket{\tilde{e}}=\ket{\tilde{0},\tilde{0}}$. This means we have to prepare the state $\ket{\widetilde{F}_{3}=1;\tilde{m}_{F}=0}$ for the three-level system and $\ket{\widetilde{F}_{2}=1/2;\tilde{m}_{F}=1/2}$ for the two level system. The preparation of the atoms representing gauge and matter fields depends on the initial physical state we want to simulate. All atoms must occupy the motional ground state with energy $E_{0}$ (different for different atomic species). As mentioned in the previous section, we also introduce a uniform magnetic field (or an AC Stark effect) to lift the degeneracy of the ground state manifolds and induce energy splittings $\Delta E$ (again different for different species) between the different $m_{F}$ components. \\
We can define the non-interacting Hamiltonian $H_{0}$ which will be present throughout the whole implementation: 

\begin{equation}
\begin{aligned}
H_{0}&=\sum_{\mathbf{x}} (E_{0,mat}+\Delta E_{mat}) \hspace{2pt} \psi_{1}^{\dagger}(\mathbf{x}) \psi_{1}(\mathbf{x})+ (E_{0,mat}-\Delta E_{mat}) \hspace{2pt} \psi_{2}^{\dagger}(\mathbf{x}) \psi_{2}(\mathbf{x}) \\
&+\sum_{\mathbf{x},k} \sum_{m_{F}} (E_{0,a}+\Delta E_{a} \hspace{2pt} m_{F}) \hspace{2pt} a^{\dagger}_{m_{F}}(\mathbf{x},k) a_{m_{F}}(\mathbf{x},k) \\
&+\sum_{\mathbf{x},k} \sum_{m_{F}} (E_{0,c}+\Delta E_{c} \hspace{2pt} m_{F}) \hspace{2pt} c^{\dagger}_{m_{F}}(\mathbf{x},k) c_{m_{F}}(\mathbf{x},k) \\
&+\sum_{\mathbf{x}} \sum_{m_{F}} (E_{0,\tilde{a}}+\Delta E_{\tilde{a}} \hspace{2pt} m_{F}) \hspace{2pt} \tilde{a}^{\dagger}_{m_{F}}(\mathbf{x}) \tilde{a}_{m_{F}}(\mathbf{x}) \\
&+\sum_{\mathbf{x}} \sum_{m_{F}} (E_{0,\tilde{c}}+\Delta E_{\tilde{c}} \hspace{2pt} m_{F}) \hspace{2pt} \tilde{c}^{\dagger}_{m_{F}}(\mathbf{x}) \tilde{c}_{m_{F}}(\mathbf{x}) \\
\label{bkg}
\end{aligned}
\end{equation}
All parts of the digital simulations are added on top of $H_{0}$. To recover the desired Hamiltonian $H$ of our $D_3$ lattice gauge theory, we move to an interaction picture , i.e. we will work in a rotating frame with respect to $H_{0}$ and make use of the rotating wave approximation. 

\subsubsection{The mass Hamiltonian}

The mass Hamiltonian in three dimensions takes the form 

\begin{equation}
H_{M}= M \sum_{\mathbf{x}} (-1)^{x_{1}+x_{2}+x_{3}} \psi^{\dagger}(\mathbf{x})\psi(\mathbf{x})
\end{equation}
with $\psi^{\dagger}(\mathbf{x})\psi(\mathbf{x})= \psi_{1}^{\dagger}(\mathbf{x})\psi_{1}(\mathbf{x})+\psi_{2}^{\dagger}(\mathbf{x})\psi_{2}(\mathbf{x})$. Thus, the corresponding time evolution $W_{M}=e^{-i H_{M}\tau}$ for a time step $\tau$ can be implemented by smoothly modulating the superlattice trapping the fermions so that the energy of the even minima is increased by an amount $M_{\mathrm{even}}$. This results in the Hamiltonian

\begin{equation}
H'_{M}= M_{\mathrm{even}} \sum_{\mathbf{x}} (1+(-1)^{x_{1}+x_{2}+x_{3}}) \psi^{\dagger}(\mathbf{x})\psi(\mathbf{x})
\end{equation}
If we act with this Hamiltonian for time $\frac{M_{\mathrm{even}}}{M} \tau$, we obtain the desired unitary evolution $W_{M}$, up to an irrelevant global phase. 

\subsubsection{Creating the isometry} \label{statorimp}
The creation of plaquette interactions and gauge-matter interactions involves constructing the isometry $S_{i}$ (see Sec. \ref{algo}), entangling auxiliary atoms with the atoms on link $i$. If we want to create it from the auxiliary state corresponding to the neutral element $\ket{\widetilde{\text{in}}}=\ket{\tilde{e}}$, we have to apply $\mathcal{U}_{i}= \int dg \ket{g}_{i} \bra{g}_{i} \otimes {\Theta^{L^{\dagger}}_{g}}$. Specifying this equation to the gauge group $D_{3}$, we obtain the following interaction between the d.o.f. on link $i$ and the ones of the control:

\begin{equation}
\begin{aligned}
\mathcal{U}_{i} \ket{\widetilde{\text{in}}}&= \sum_{p} \sum_{m} \ket{p,m}_{i} \bra{p,m}_{i} \otimes {\Theta^{L^{\dagger}}_{p,m}} \ket{\tilde{0},\tilde{0}} \\
&=\sum_{p} \sum_{m} \ket{p,m}_{i} \bra{p,m}_{i} \otimes \widetilde{Q}_{3}^{p} \widetilde{Q}_{2}^{m} \ket{\tilde{0},\tilde{0}}=\widetilde{Q}_{3,i}^{\hat{p}} \widetilde{Q}_{2,i}^{\hat{m}} \ket{\tilde{0},\tilde{0}}
\end{aligned}
\end{equation}
where $\hat p = \sum_p p \ket{p}_{i}\bra{p}_{i}$ and $\hat m = \sum_m m \ket{m}_{i}\bra{m}_{i}$. As defined previously, $\widetilde{Q}_{2}$ and $\widetilde{Q}_{3}$ are the raising operators of the auxiliary atoms, i.e $\hspace{2pt} \widetilde{Q}_{2,i}^{\hat{m}}$ and $\widetilde{Q}_{3,i}^{\hat{p}}$ raise the $\tilde{m}_{F}$-values of the auxiliary atoms according to the $m_{F}$-values of the atoms on link $i$. By choosing $\ket{\tilde{0},\tilde{0}}$ as the initial state of the auxiliary atoms, the creation of the isometry reduces to an interaction between the three-level atom on the link and the auxiliary three-level atom in parallel with an interaction between the two-level atom on the link and the auxiliary two-level atom. These are the same interactions required for creating the isometry of a $\mathbb{Z}_{3}$ lattice gauge theory \cite{zohar2017digital}, respectively a $\mathbb{Z}_{2}$ lattice gauge theory \cite{zohar2016digital}. We can therefore directly adopt the procedure from \cite{zohar2017digital,zohar2016digital}. The idea is to write $\widetilde{Q}_{3,i}^{\hat{p}}$ and $\widetilde{Q}_{2,i}^{\hat{m}}$ as an interaction between the z-components of the hyperfine angular momentum operators $\widetilde{F}_{z,3}$ and $F_{z,3}$, respectively $\widetilde{F}_{z,2}$ and $F_{z,2}$: 

\begin{equation}
\widetilde{Q}_{3,i}^{\hat{p}}=\widetilde{V}_{3}^{\dagger}\mathcal{U}'_{3,i}\widetilde{V}_{3} \hspace{20pt} \mathrm{with} \hspace{10pt} \mathcal{U}'_{3,i}= e^{i \frac{2\pi}{3} \widetilde{F}_{z,3} F_{z,3}}
\end{equation}
where $\widetilde{V}_{3}^{\dagger}$ is a local change of basis from the $\widetilde{P}_{3}$-basis $\{ \ket{\tilde{p}} \}$ to its conjugate $\widetilde{Q}_{3}$-basis and:

\begin{equation}
\widetilde{Q}_{2,i}^{\hat{m}}=\widetilde{V}_{2}^{\dagger}\mathcal{U}'_{2,i}\widetilde{V}_{2} \hspace{20pt} \mathrm{with} \hspace{10pt} \mathcal{U}'_{2,i}=e^{-i \frac{1}{\pi} (i\frac{\pi}{2})^2  (1-2 F_{z,2})  (1-2 \widetilde{F}_{z,2})}
\end{equation}
where $\widetilde{V}_{2}^{\dagger}$ is mapping from the $\widetilde{P}_{2}$-basis $\{ \ket{\tilde{m}} \}$ into the conjugate $\widetilde{Q}_{2}$-basis. The basis transformations $\widetilde{V}_{3}$ and $\widetilde{V}_{2}$ are local operations on the auxiliary atoms that can be implemented with optical/RF fields. The interactions between the z-components of the hyperfine angular momentum operator can be realized by introducing an energy splitting between the different $m_{F}$-levels such that the two-body scattering term will contain only $m_{F}$ preserving terms. The sequence to obtain $\mathcal{U}_{i}$ is therefore: 
\begin{equation}
\mathcal{U}_{i}=\widetilde{V}_{3}^{\dagger}\widetilde{V}_{2}^{\dagger}\mathcal{U}'_{3,i}\mathcal{U}'_{2,i}\widetilde{V}_{2}\widetilde{V}_{3}
\end{equation}
To undo the isometry it is necessary to create the conjugate of these interactions which can be done by flipping locally the $\tilde{m}_{F}=1$ and $\tilde{m}_{F}=-1$ levels, thus mapping $\widetilde{F}_{z,3}$ into $-\widetilde{F}_{z,3}$. In the same way, the conjugate of the two-level system is created.

\subsubsection{Plaquette interactions}
Knowing how to construct the isometry, the implementation of the plaquette interactions is straightforward. Since we have to split them in six different parts (see Sec. \ref{algo}), we start with $H_{B,1e}$, the type 1 plaquettes of the even cubes, where the auxiliary atoms are placed in the standard configuration. We follow the three steps of the algorithm given in Sec. \ref{algoplaq}: 

\begin{enumerate}
\item{We create the plaquette isometry out of the isometries $S_{i}$ which is realized for a link $i$ by moving the lattice of the auxiliary atoms to the respective link and tailoring the interactions as described above (neglecting the basis transformations $\widetilde{V}$ for the moment). This can be done in parallel for the whole lattice:

\begin{equation} \label{groupimplement}
\mathcal{U}_{ie}^{\prime}= \prod_{\mathbf{x} \hspace{2pt} \mathrm{even}} \mathcal{U}'_{3, i} (\mathbf{x}) \hspace{2pt} \mathcal{U}'_{2,  i} (\mathbf{x}) 
\end{equation}
The desired plaquette isometry is obtained by applying this procedure to all four links and including overall basis transformations $\widetilde{V}_{3,all}$ and $\widetilde{V}_{2,all}$:

\begin{equation}
\prod\limits_{\mathbf{x} \hspace{2pt} \mathrm{even}} \mathcal{U}_{1} (\mathbf{x})\mathcal{U}_{2}(\mathbf{x}) \mathcal{U}_{3}^{\dagger} (\mathbf{x}) \mathcal{U}_{4}^{\dagger} (\mathbf{x}) = \widetilde{V}^{\dagger}_{3,all}\widetilde{V}^{\dagger}_{2,all} \mathcal{U}'_{1e} \mathcal{U}'_{2e} \mathcal{U}^{\prime \dagger}_{3e} \mathcal{U}^{\prime \dagger}_{4e} 
\widetilde{V}_{2,all} \widetilde{V}_{3,all}
\end{equation}
This operation, acting on the tensor product of $\ket{\tilde{0},\tilde{0}}$ and any state of the links, gives rise to the proper entangled state which maps plaquette interactions to local operations on the control. 
}

\item{The next step is a local operation on the auxiliary Hilbert space. We need to implement $e^{-i\widetilde{H}_{B}\tau}$ with $\widetilde{H}_{B} $ being the control Hamiltonian $\widetilde{H}_{B}=\lambda_{B} \mathrm{Tr} \hspace{1pt} (\widetilde{U}+\widetilde{U}^{\dagger})$. This requires an interaction between the two-level and the three-level system:

\begin{equation}
\begin{aligned}
\widetilde{H}_{B}&=\lambda_{B} \hspace{1pt} \mathrm{Tr} \left(\sum_{p} \sum_{m} \ket{\tilde{p},\tilde{m}} \bra{\tilde{p},\tilde{m}} \hspace{2pt} e^{i\frac{2\pi}{3}\sigma_{z}p} \sigma_{x}^{m} +H.c. \right) \\
&=2\lambda_{B} \hspace{1pt} (\widetilde{P}_{3}+\widetilde{P}_{3}^{\dagger}) (1-\hat{m})
\end{aligned}
\end{equation}
where $\hat m = \sum_m m \ket{\tilde{m}}\bra{\tilde{m}}$. We can rewrite $\widetilde{P}_{3}+\widetilde{P}_{3}^{\dagger} = -\mathbb{I} + 3 \ket{\tilde 0}\bra{\tilde 0} \equiv -\mathbb{I} + 3 N_0$ with $\widetilde{N}_{0} \equiv \tilde{a}_{0}^{\dagger} \tilde{a}_{0}$}. Defining a number operator for the $\ket{\widetilde{F}_{2}=1/2; m_{F}=1/2}$ state of the two-level system as $\widetilde{N}_{1/2} \equiv \tilde{c}_{1/2}^{\dagger} \tilde{c}_{1/2}$ we can write down the interaction $e^{-i\widetilde{H}_{B}\tau}$:

\begin{equation}
\begin{aligned}
e^{-i\widetilde{H}_{B}\tau}=e^{-i2\lambda_{B} \hspace{1pt} (-\mathbb{I}+3 \widetilde{N}_{0}) \widetilde{N}_{1/2}  \tau }=e^{i2\lambda_{B} \widetilde{N}_{1/2} \tau} \hspace{1pt} e^{-i6\lambda_{B}\widetilde{N}_{0}\widetilde{N}_{1/2} \tau}
\end{aligned}
\end{equation}
The first exponential is a local term of the two-level system which can be implemented by means of optical/RF fields. The second term requires scattering between the two auxiliary atoms. The corresponding Hamiltonian density in second quantized form is \cite{lewenstein2012ultracold}: 

\begin{equation}
\mathcal{H}_{scat} (\mathbf{x}) =\frac{2 \pi}{\mu}  \sum_{\alpha,\beta,\gamma,\delta} \sum_{k=0}^{n-1}  g_{k} ((\mathbf{F}_{1} \cdot \mathbf{F}_{2})^k)_{\alpha,\beta,\gamma,\delta} \Phi^{\dagger}_{\alpha} (\mathbf{x}) \Phi^{\dagger}_{\beta} (\mathbf{x}) \Phi_{\gamma} (\mathbf{x}) \Phi_{\delta} (\mathbf{x})
\label{EqVB}
\end{equation}
where $\Phi^{\dagger}_{\alpha}$ denotes the creation operator of the atomic Wannier wave function corresponding to the internal state $\alpha$ and $\mu$ the reduced mass of the two atomic species. The projection operators onto the different scattering channels are expressed by polynomials of $\mathbf{F}_{1} \cdot \mathbf{F}_{2}$, the coefficients $g_{k}$ are therefore functions of the scattering lengths. To obtain the time evolution due to this interaction we have to integrate the Hamiltonian density over space and time. Since eq. \eqref{EqVB} involves only specific levels, we need to turn on the magnetic field gradient and split the different $m_F$ components such that only the $\tilde{m}_{F}=0$-component and the $\tilde{m}_{F}=1/2$-component overlap during the collision. Moreover, changes in the internal level of the two atoms during the collision are suppressed by the Zeeman splitting. With these assumptions, the time evolution is described by the following unitary

\begin{equation}
\mathcal{U}_{scat,1}=\mathbb{I} + (e^{-i g_0 \alpha }-1) \ket{\tilde 0, \tilde{\tfrac{1}{2}}} \bra{\tilde 0, \tilde{\tfrac{1}{2}}} = e^{-i g_0 \alpha \widetilde{N}_{0} \widetilde{N}_{1/2}}
\end{equation}
with $g_{0}=\frac{1}{6} (3a_{1/2} + 4a_{3/2})$ ($a_{1/2},a_{3/2}$ are the scattering lengths for the scattering channels with $F_{tot}=1/2$ and $F_{tot}=3/2$) and $\alpha$ the time-integral of the overlap of the two wave-functions during the collision. By carefully tuning the interaction time we can set $\alpha=\frac{6\lambda_{B}\tau}{g_{0}}$ and finally obtain: 

\begin{equation}
\mathcal{U}_{scat,1}=e^{-i 6\lambda_{B}  \widetilde{N}_{0} \widetilde{N}_{1/2} \tau}
\end{equation}
which is up to local operations the desired unitary $V_{B}$. This interaction will be implemented in parallel for all cubes where auxiliary atoms are placed, i.e. in this case for the even cubes. Hence, the overall interaction of this step is $e^{-i\sum\limits_{\mathbf{x} \hspace{2pt} \mathrm{even}} \widetilde{H}_{B} (\mathbf{x})\tau}$. \\
When the magnetic field gradient is on, different levels of the hyperfine multiplet will acquire an extra energy splitting with respect to the background Hamiltonian \eqref{bkg}. This induces extra phases that need to be cancelled somehow. For example, after the collision has been completed, we can invert the slope of the gradient and accumulate phases in the opposite direction until the net effect is zero (this trick has to be applied for all scattering events of this kind).

\item{In the third and last step we have to undo the isometry. This can be done by taking the hermitian conjugate of the first step, i.e. the sequence: 

\begin{equation}
\widetilde{V}^{\dagger}_{3,all}\widetilde{V}^{\dagger}_{2,all} \mathcal{U}'_{4e} \mathcal{U}'_{3e} \mathcal{U}^{\prime \dagger}_{2e} \mathcal{U}^{\prime \dagger}_{1e} 
\widetilde{V}_{2,all} \widetilde{V}_{3,all}
\end{equation}

}

\end{enumerate}
According to ($\ref{timeevmag}$) these three steps give us $W_{B,1e}$. If we repeat now the same procedure but with the links corresponding to the second and third plaquette term, we obtain $W_{B,2e}$ and $W_{B,3e}$. To realize the odd cubes time evolution, we move the auxiliary atoms to the centers of the odd cubes and repeat all of the above. This results in the time evolutions $W_{B,1o}$,$W_{B,2o}$ and $W_{B,3o}$. Afterwards, the auxiliary atoms are brought back to the centers of the even cubes.

\subsubsection{Gauge-matter interactions}

For the Gauge-matter interactions on a link $(\mathbf{x},k)$ we have to implement the Hamiltonian
\begin{equation}
\begin{aligned}
H_{GM}(\mathbf{x},k) &=\lambda_{GM} \hspace{1pt} \psi_{a}^{\dagger}(\mathbf{x}) U_{ab}(\mathbf{x},k) \psi_{b}(\mathbf{x}+\mathbf{k}) + H.c.  \\
&=\lambda_{GM} \hspace{1pt} \psi_{a}^{\dagger}(\mathbf{x})  \hspace{2pt} (e^{i\frac{2\pi}{3}\sigma_{z}\hat{p}})_{ab} \hspace{1pt} (\sigma_{x}^{\hat{m}})_{bc} \hspace{2pt} \psi_{c}(\mathbf{x}+\mathbf{k}) + H.c. \\
&=\lambda_{GM} \hspace{1pt} \psi_{a}^{\dagger}(\mathbf{x})  \hspace{2pt} (U_{p})_{ab}(\mathbf{x},k) \hspace{1pt} (U_{m})_{bc}(\mathbf{x},k) \hspace{2pt} \psi_{c}(\mathbf{x}+\mathbf{k}) + H.c.
\end{aligned}
\end{equation}
with $U_{p} \equiv e^{i\frac{2\pi}{3}\sigma_{z}\hat{p}}$ and $U_{m} \equiv \sigma_{x}^{\hat{m}}$. We can use the product structure of $U$ to implement the gauge-matter part via two-body interactions. We follow the procedure given in Sec. \ref{algogauge} and define the unitaries $\mathcal{U}_{W}$, one corresponding to $U_{p}$: 
\begin{equation}
\mathcal{U}_{W,p}(\mathbf{x},k)=e^{\log(U_p)_{ab}(\mathbf{x},k) \hspace{1pt} \psi_{a}^{\dagger}(\mathbf{x})\psi_{b}(\mathbf{x})}=e^{i\frac{2\pi}{3} \hat{p} \hspace{1pt} (\psi_{1}^{\dagger}(\mathbf{x})\psi_{1}(\mathbf{x})-\psi_{2}^{\dagger}(\mathbf{x})\psi_{2}(\mathbf{x}))}
\end{equation}
and another one corresponding to $U_{m}$:
\begin{equation}
\mathcal{U}_{W,m}(\mathbf{x},k)=e^{\log(U_m)_{ab}(\mathbf{x},k) \hspace{1pt} \psi_{a}^{\dagger}(\mathbf{x})\psi_{b}(\mathbf{x})}=e^{i\frac{\pi}{2}\hat{m} \hspace{1pt} (\psi_{1}^{\dagger}(\mathbf{x})\psi_{1}(\mathbf{x})+\psi_{2}^{\dagger}(\mathbf{x})\psi_{2}(\mathbf{x})-\psi_{1}^{\dagger}(\mathbf{x})\psi_{2}(\mathbf{x})-\psi_{2}^{\dagger}(\mathbf{x})\psi_{1}(\mathbf{x}))}
\end{equation}
With these definitions at hand we can get the following relation by applying twice the Baker-Campbell-Hausdorff formula:
\begin{equation}
\mathcal{U}_{W,p}(\mathbf{x},k)\mathcal{U}_{W,m}(\mathbf{x},k) \psi_{n}^{\dagger}(\mathbf{x})\mathcal{U}_{W,m}^{\dagger}(\mathbf{x},k)\mathcal{U}_{W,p}^{\dagger}(\mathbf{x},k)=
\psi_{a}^{\dagger}(\mathbf{x})(U_p)_{ab}(\mathbf{x},k)(U_m)_{bn}(\mathbf{x},k)
\end{equation}
The gauge-matter Hamiltonian can then be written as
\begin{equation}
H_{GM}(\mathbf{x},k)=\mathcal{U}_{W,p}(\mathbf{x},k)\mathcal{U}_{W,m}(\mathbf{x},k) H_{t}(\mathbf{x},k) \mathcal{U}_{W,m}^{\dagger}(\mathbf{x},k)\mathcal{U}_{W,p}^{\dagger}(\mathbf{x},k)
\end{equation}
with the tunneling Hamiltonian $H_{t}(\mathbf{x},k)=\lambda_{GM} \hspace{1pt} (\psi_{a}^{\dagger}(\mathbf{x}) \psi_{a}(\mathbf{x}+\mathbf{k}) + H.c.)$ The crucial thing to note here is that all the terms involve only two-body interactions which allows an implementation with the proposed ultracold atomic setup. We can not implement all gauge-matter interactions at once as the fermions on the vertices are only allowed to interact with one link at a time. Focusing on the links in the $\mathbf{1}$-direction for the even cubes, we describe how to realize the time evolution $e^{-i \sum\limits_{\mathbf{x} \hspace{2pt} \mathrm{even}} H_{GM}(\mathbf{x},1) \tau}$. Since we want to keep the lattice of the matter and link degrees of freedom fixed, these interactions will be mediated by the control atoms according to the algorithm presented in \ref{algogauge}. 
\begin{enumerate}
\item{We first build the isometry $S_{1}$ between auxiliary atoms located at the center of even cubes $\mathbf{x}$ and the corresponding atoms on link $(\mathbf{x},1)$, $\prod\limits_{\mathbf{x} \hspace{2pt} \mathrm{even}} \mathcal{U}_{1} (\mathbf{x})$. This interaction can be implemented exactly in the same way as already done for the plaquette term (see (\ref{groupimplement})). Due to the relation in (\ref{eigenoperatorgroup}) the gauge-matter interactions will then translate into an interaction of exactly the same form but between the auxiliary atoms and the fermions.}

\item{Afterwards, the two terms $\mathcal{U}_{W,p}^{\dagger}$ and $\mathcal{U}_{W,m}^{\dagger}$ have to be implemented by two-body scattering processes but between the fermions and the auxiliary atoms due to the isometry, therefore denoted as $\widetilde{\mathcal{U}}_{W,p}$ and $\widetilde{\mathcal{U}}_{W,m}$. Starting with $\widetilde{\mathcal{U}}_{W,p}^{\dagger}$, we first write it in terms of the angular momentum operator respectively the second quantized operators $\psi_{1}$ and $\psi_{2}$ for the fermions: 
\begin{equation}
\widetilde{\mathcal{U}}_{W,p}^{\dagger}(\mathbf{x},k)=e^{-i\frac{2\pi}{3} \hat{p} \hspace{1pt} (\psi_{1}^{\dagger}(\mathbf{x})\psi_{1}(\mathbf{x})-\psi_{2}^{\dagger}(\mathbf{x})\psi_{2}(\mathbf{x}))}= e^{-i\frac{2\pi}{3} \widetilde{F}_{z,3} \hspace{1pt} (\psi_{1}^{\dagger}(\mathbf{x})\psi_{1}(\mathbf{x})-\psi_{2}^{\dagger}(\mathbf{x})\psi_{2}(\mathbf{x}))}
\end{equation}
Now we have to tailor the atomic collision between the $\widetilde{F}_{3}=1$ and the $F=1/2$ multiplet accordingly. The magnetic field again lifts the degeneracy of the hyperfine levels and thereby prevents any transitions changing the $m_{F}$-values. The interaction Hamiltonian contains two possible scattering channels and gives rise to the following time evolution:
\begin{equation}
\mathcal{U}_{scat,2}=e^{-i \beta (g_{0} (\psi_{1}^{\dagger}\psi_{1}+\psi_{2}^{\dagger}\psi_{2}) +g_{1} \widetilde{F}_{z,3} (\psi_{1}^{\dagger}\psi_{1}-\psi_{2}^{\dagger}\psi_{2}))}
\end{equation}
with $g_{0}=\frac{1}{6} (3a_{1/2} + 4a_{3/2})$, $g_{1}=\frac{2}{3} (a_{3/2} - a_{1/2})$ ($a_{1/2},a_{3/2}$ are the scattering lengths for the scattering channels with $F_{tot}=1/2$ and $F_{tot}=3/2$) and $\beta$ the time-integral of the wave-function overlap. If we tune overlap and interaction time such that $\beta=\frac{2 \pi}{3 g_{1}}$ we obtain 
\begin{equation}
\mathcal{U}_{scat,2}=e^{-i \frac{2 \pi g_{0}} {3 g_{1}} (\psi_{1}^{\dagger}\psi_{1}+\psi_{2}^{\dagger}\psi_{2})} e^{-i\frac{2 \pi}{3} \widetilde{F}_{z,3} (\psi_{1}^{\dagger}\psi_{1}-\psi_{2}^{\dagger}\psi_{2})}
\end{equation}
The second exponential is the desired interaction $\widetilde{\mathcal{U}}_{W,p}^{\dagger}$ whereas the first term is a fermion-dependent phase, denoted from now on as 
\begin{equation}
V_{W'}(\theta)=e^{-i\theta \psi^{\dagger} \psi}
\end{equation}
where $\theta=\frac{2 \pi g_{0}} {3 g_{1}}$ and $\psi^{\dagger} \psi=\psi_{1}^{\dagger}\psi_{1}+\psi_{2}^{\dagger}\psi_{2}$. A discussion of these phases will be done later on. Before, the implementation of $\widetilde{\mathcal{U}}_{W,m}^{\dagger}$ is explained. It has the form: 

\begin{equation}
\begin{aligned}
\widetilde{\mathcal{U}}_{W,m}^{\dagger}&=e^{-i\frac{\pi}{2}\hat{m} \hspace{1pt} (\psi_{1}^{\dagger}\psi_{1}+\psi_{2}^{\dagger}\psi_{2}-\psi_{1}^{\dagger}\psi_{2}-\psi_{2}^{\dagger}\psi_{1})} \\
&=e^{-i\frac{\pi}{2} \widetilde{N}_{-1/2} \hspace{1pt} (\psi_{1}^{\dagger}\psi_{1}+\psi_{2}^{\dagger}\psi_{2}-\psi_{1}^{\dagger}\psi_{2}-\psi_{2}^{\dagger}\psi_{1})}=V_{H,fer} e^{-i \pi \widetilde{N}_{-1/2} \hspace{1pt} \psi_{2}^{\dagger}\psi_{2}} V_{H,fer}
\end{aligned}
\end{equation}
with $\widetilde{N}_{-1/2}\equiv \tilde{c}_{-1/2}^{\dagger} \tilde{c}_{-1/2}$ and $V_{H,fer}=\frac{1}{\sqrt[]{2}} (\sigma_{x,fer}+\sigma_{z,fer})$ a Hadamard transform on the fermions which can be implemented by means of optical/RF fields. The remaining two-body interaction is realized as scattering between the $F=1/2$ states of the control atoms and the fermions. It can be described by the following unitary: 
\begin{equation}
\mathcal{U}_{scat,3}=e^{-i\gamma \left(g_{0} \sum_{m} \tilde{c}_{m}^{\dagger} \tilde{c}_{m} (\psi_{1}^{\dagger}\psi_{1}+\psi_{2}^{\dagger}\psi_{2}) + g_{1} \widetilde{F}_{z,2} (\psi_{1}^{\dagger}\psi_{1}-\psi_{2}^{\dagger}\psi_{2}) \right)}
\end{equation}
(for the explicit form of $g_{k}$ see \cite{zohar2016digital}). We switch on a magnetic field gradient designed in a way that only the $m_{F}=-1/2$ -components of the auxiliary atom and the fermion overlap. Moreover, the interaction time should be tuned such that $\gamma=\frac{\pi}{g_{0}+g_{1}}$ which gives rise to: 
\begin{equation}
\mathcal{U}_{scat,3}=e^{-i \gamma \left(g_{0} \tilde{c}_{-1/2}^{\dagger} \tilde{c}_{-1/2} \psi_{2}^{\dagger}\psi_{2} + g_{1} \tilde{c}_{-1/2}^{\dagger} \tilde{c}_{-1/2} \psi_{2}^{\dagger}\psi_{2}  \right)}=e^{-i \pi \tilde{c}_{-1/2}^{\dagger} \tilde{c}_{-1/2} \hspace{1pt} \psi_{2}^{\dagger}\psi_{2}}
\end{equation}
Since the implementation of $\widetilde{\mathcal{U}}_{W,p}^{\dagger}$ and $\widetilde{\mathcal{U}}_{W,m}^{\dagger}$ is done in parallel for all even cubes we get the sequence
\begin{equation}
\prod\limits_{\mathbf{x} \hspace{2pt} \mathrm{even}} \widetilde{\mathcal{U}}_{W,m}^{\dagger}(\mathbf{x},1) \widetilde{\mathcal{U}}_{W,p}^{\dagger}(\mathbf{x},1)  V_{W'}(\theta)
\end{equation}}

\item{In the next step we implement the tunneling in the 1-direction for even cubes which can be achieved by modulating the superlattice and decreasing the potential barriers on the desired links. We get 
\begin{equation}
\prod_{\mathbf{x} \hspace{2pt} \mathrm{even}} e^{-i H_{t}(\mathbf{x},1) \tau}
\end{equation}}

\item{After the tunneling we need to realize the conjugate of $\widetilde{\mathcal{U}}_{W,p}^{\dagger}$ and $\widetilde{\mathcal{U}}_{W,m}^{\dagger}$, i.e. $\widetilde{\mathcal{U}}_{W,p}$ and $\widetilde{\mathcal{U}}_{W,m}$. One way of creating  $\widetilde{\mathcal{U}}_{W,p}$ is by doing a spin flipping operation $\widetilde{V}_{F,3}$ for the three-level system of the control which results in: 
\begin{equation}
\begin{aligned}
\widetilde{V}_{F,3} \widetilde{\mathcal{U}}_{W,p}^{\dagger} \widetilde{V}_{F,3}^{\dagger}&=\widetilde{\mathcal{U}}_{W,p} \\
\widetilde{V}_{F,3} V_{W'}(\theta) \widetilde{V}_{F,3}^{\dagger}&=V_{W'}(\theta)
\end{aligned}
\end{equation}
For the creation of $\widetilde{\mathcal{U}}_{W,m}$ we simply observe that $\widetilde{\mathcal{U}}_{W,m}^{\dagger}$ is hermitian. The sequence for step 4 is
\begin{equation}
\prod_{\mathbf{x} \hspace{2pt} \mathrm{even}} V_{W'}(\theta) \widetilde{\mathcal{U}}_{W,p}(\mathbf{x},1) \widetilde{\mathcal{U}}_{W,m}(\mathbf{x},1)    
\end{equation}}

\item{In the last step we need to undo the isometry, which is done by the conjugate of the first step, $\prod\limits_{\mathbf{x} \hspace{2pt} \mathrm{even}} \mathcal{U}_{1}^{\dagger} (\mathbf{x})$ (see Sec. \ref{statorimp}). }
\end{enumerate}
We summarize by writing down the whole sequence acting on the initial auxiliary state $\ket{\widetilde{\text{in}}}= \ket{\tilde{0},\tilde{0}}$:
\begin{equation}
\begin{aligned}
&\prod_{\mathbf{x} \hspace{2pt} \mathrm{even}} V_{W'}(\theta) \mathcal{U}_{1}^{\dagger}(\mathbf{x}) \widetilde{\mathcal{U}}_{W,p}(\mathbf{x},1) \widetilde{\mathcal{U}}_{W,m}(\mathbf{x},1) e^{-i H_{t}(\mathbf{x},1) \tau}  \widetilde{\mathcal{U}}_{W,m}^{\dagger}(\mathbf{x},1) \widetilde{\mathcal{U}}_{W,p}^{\dagger}(\mathbf{x},1)  \mathcal{U}_{1} (\mathbf{x}) V_{W'}(\theta)   \ket{\widetilde{\text{in}}}  \\
=& \ket{\widetilde{\text{in}}} V_{W'}(\theta) \prod_{\mathbf{x} \hspace{2pt} \mathrm{even}} e^{-i H_{GM}(\mathbf{x},1) \tau} V_{W'}(\theta)
=\ket{\widetilde{\text{in}}} V_{W'}(\theta) W_{GM,1e} V_{W'}(\theta)
\label{seq92}
\end{aligned}
\end{equation}
We finally get the desired gauge-matter interactions up to the fermionic phases $V_{W'}(\theta)$. However, if we consider the whole lattice (on which the number of fermions is globally conserved) it can be shown that the phases correspond to a static vector potential of zero magnetic field and are therefore unphysical, as carried out in the procedure given in \cite{zohar2017digital}. If we repeat the whole sequence \eqref{seq92} for the other links we obtain the gauge-matter interactions $W_{GM,2e}$, $W_{GM,3e}$, $W_{GM,1o}$, $W_{GM,2o}$ and $W_{GM,3o}$.

\subsubsection{Electric Hamiltonian}
The electric Hamiltonian for the gauge group $D_{3}$ acts on the gauge fields residing on the links. If we choose its second part - which corresponds to pure rotations only - in accordance with the electric energy of $\mathbb{Z}_{3}$ we obtain, using the notation of previous sections:
\begin{equation}
\begin{aligned}
H_{E}&=\lambda_{E} \sum_{\mathbf{x},k} h_{E}(\mathbf{x},k) \\
\mathrm{with} \hspace{20pt} h_{E}(\mathbf{x},k)&=\frac{1}{2} f_{r} \sum_{m,m'} \ket{0,m} \bra{0,m'} + f_{l}  (1-P_{3}-P_{3}^{\dagger}) \otimes \mathbb{I}_{2}
\end{aligned}
\end{equation}
If we also express the interactions of the first part in terms of operators acting on the link atoms, we end up with: 
\begin{equation}
\begin{aligned}
h_{E}(\mathbf{x},k)=\frac{1}{2} f_{r} a_{0}^{\dagger}a_{0} \otimes (1+\sigma_{x}) + f_{l} \sum_{m_{F}=-1}^{1} (1+|m_{F}|) a_{m_{F}}^{\dagger}a_{m_{F}}  \otimes \mathbb{I}_{2}
\end{aligned}
\end{equation}
The first Hilbert space represents the three-level system, the second one the two-level system. The coefficient $f_{l}$ is the overall coefficient for the electric part corresponding to pure rotations, equivalently to $\mathbb{Z}_{3}$. We have to implement the time evolution: 
\begin{equation}
W_{E}=e^{-i H_{E} \tau}= \prod_{\mathbf{x},k} e^{-i h_{E}(\mathbf{x},k) \tau}
\end{equation}
with 
\begin{equation}
e^{-i h_{E} \tau}=e^{-i \frac{\lambda_{E}f_{r}}{2} a_{0}^{\dagger}a_{0} \tau} e^{-i \frac{\lambda_{E}f_{r}}{2} a_{0}^{\dagger}a_{0} \sigma_{x} \tau} e^{-i \lambda_{E} f_{l} \sum_{m_{F}} (1+|m_{F}|) a_{m_{F}}^{\dagger}a_{m_{F}}  \tau}
\end{equation}
The first and the third exponential are local terms of the atoms and can be addressed by external fields. The second term is implemented by two-body scattering similar to the one for the plaquette interactions. Therefore, we need to bring the two atoms together, which should be simple to implement since both of them are trapped near the middle of the link. Following the steps for the plaquette interactions, we obtain: 

\begin{equation}
\mathcal{U}_{scat,1}=e^{-i\delta g_{0} N_{0} N_{1/2}} 
\end{equation}
Tuning overlap and interaction time such that $\delta=\frac{\lambda_{E} f_{r} \tau}{g_{0}}$ and combining it with the local operation $\mathcal{V}_{2}=e^{i \frac{\lambda_{E} f_{r} \tau}{2} N_{1/2}}$, gives us: 
\begin{equation}
\mathcal{V}_{2}\mathcal{U}_{scat,1}=e^{i \frac{\lambda_{E} f_{r} \tau}{2} N_{0}} e^{-i \lambda_{E} f_{r} N_{0} N_{1/2} \tau}=e^{-i \frac{\lambda_{E} f_{r}}{2} N_{0} \sigma_{z} \tau}
\end{equation}
If we then perform a Hadamard transform $V_{H,2}$ on the two-level system, we get the desired interaction: 
\begin{equation}
V_{H,2}\mathcal{V}_{2}\mathcal{U}_{scat,1}V_{H,2}=e^{-i \frac{\lambda_{E} f_{r}}{2} N_{0} \sigma_{x}}=e^{-i \frac{\lambda_{E}f_{r}}{2} a_{0}^{\dagger}a_{0} \sigma_{x} \tau}
\end{equation}
which gives us the electric Hamiltonian up to local operations. \\
We have implemented all interactions using local operations on the atoms and tailoring the appropriate two-body scattering terms. If we use the sequence to evolve the system for a time $\tau = T/N$ and we repeat the same sequence $N$ times, we get a Trotter approximation of the desired time-evolution $e^{-i H_{LGT} T}$. The accuracy of this approximation is discussed below.

\subsubsection{Errors}
The errors affecting the precision of the simulation are twofold. On the one hand, we have Trotter errors coming from the digitization which can be estimated by specifying the general error bounds given in Sec. \ref{algo} to the case of three dimension and gauge group $D_{3}$. We obtain for the fist order formula (see (\ref{firstordertrott})): 
\begin{equation}
\begin{aligned}
\|\mathcal{U}(t)-\mathcal{U}_{N}(t) \| \le\frac{6 t^2 (L-1)L^2}{N} \left( 16 \lambda_{B} \lambda_{E} + 2 \lambda_{GM}  \lambda_{E}  + M \lambda_{GM} + \lambda_{GM}^2 \frac{5}{4} \right) 
\end{aligned}
\end{equation}
and the second order formula (see (\ref{secondordertrott})): 
\begin{equation}
\begin{aligned}
\|\mathcal{U}(t)-\mathcal{U}_{2,N}(t) \| \le& \frac{2t^3  (L-1) L^{2}}{ N^2} \left(128 \lambda_{E}\lambda_{B}  ( \lambda_{E}  + \lambda_{B}    ) + 2\lambda_{GM}\lambda_{E} ( 22 \lambda_{GM}  + \lambda_{E}    ) \right.\\
&+ \left. \lambda_{GM}M ( 6 \lambda_{GM}+\frac{1}{2} M ) +  \frac{125}{12} \lambda_{GM}^3     \right)
\end{aligned}
\end{equation}
We stress again that the digitization error doesn't break gauge invariance, because all steps of the sequence individually respect the right symmetry. Therefore, the Trotter expansion can only give rise to quantitative deviations, but not to qualitative changes. \\
On the other hand, there will be experimental errors in the implementation. Unlike errors caused by the Trotterization, they may break the gauge symmetry and accumulate step by step. We briefly want to look at the scaling of these errors. We consider a small perturbation $h_{j}$ to one of the Hamiltonians $H_{j}$ which is realized during the implementation of the Trotter sequence (\ref{sequence}). The difference of the time evolution $e^{-i \tau (H_{j}+h_{j})}$ to the desired one $e^{-i\tau H_{j}}$ can be bounded to first order in the operator norm by $ \| h_{j} \| \tau$. To get the total experimental error caused by the gates corresponding to $H_{j}$, we need to look at the whole Trotterized time evolution (we focus here on the second order formula (\ref{ZweiteOrdnung}), i.e. the gate is repeated $2N$ times). We have to distinguish four cases: On the one hand, whether the experimental error is statistical or systematic and, on the other hand, whether the implemented gate depends on the simulated time (e.g. electric Hamiltonian, fermionic tunneling, etc.) or not (e.g. entangling operations). The advantage of a statistical error is that we can apply the central limit theorem and obtain a scaling of $\sqrt[]{N}$ with the number of Trotter steps compared to a linear scaling in the case of a systematic error. In the same vein, a gate depending on the simulated time $t$ is advantageous since the time step $\tau=\frac{t}{2N}$ in each trotter sequence scales as $\frac{1}{N}$, whereas for gates not depending on $t$, the error scales with some fixed amount of time $t_{\text{exp},j}$ specific to the gate. The bounds for these four types of experimental errors are summarized in Table \ref{table:errors}. We see that operations that do not depend on the simulated time $t$ are the ones most prone to errors. During their implementation a lot of care should be taken, in particular to avoid systematic errors. When estimating the error of the whole implementation sequence, one should keep in mind that errors of different gates are generally independent and thus do not add up linearly. However, the total experimental error will still increase with $N$, so that the number of Trotter steps has to be chosen in a way to balance digitization and implementation errors.

\begin{table} 
\centering
\begin{tabular} {|c|c|}
\hline
Type of error & Bound on the experimental error \\ \hline
Statistical error/dependence on $t$ & $\| h_{j} \| \frac{t}{\sqrt[]{2N}}$ \\ \hline
Systematic error/dependence on $t$ & $\| h_{j} \| t$ \\ \hline
Statistical error/no dependence on $t$ & $\| h_{j} \| \hspace{2pt} \sqrt[]{2N} \hspace{2pt} t_{\text{exp},j}$ \\ \hline
Systematic error/no dependence on $t$ & $\| h_{j} \| 2N \hspace{2pt} t_{\text{exp},j} $ \\ 
\hline
\end{tabular} 
\caption{The different types of experimental errors corresponding to some gate $j$ are distinguished by the nature of the perturbation $h_{j}$ (statistical or systematic) and whether the gate depends on the simulated time $t$ or is a fixed operation lasting for some time $t_{\text{exp},j}$. The error bound for each type scales differently with the number of Trotter steps $N$.}
\label{table:errors}
\end{table} 
Typical sources of errors in ultracold atom experiments are as follows: The first one is decoherence, e.g. caused by spontaneous scattering of lattice photons with the atoms, atomic collisions with the background gas, field (laser or magnetic) fluctuations, etc. This is relatively well under control nowadays, where coherence times $t_{\text{coh}}$ of the order of minutes have already been achieved \cite{hamann1998resolved,friebel1998co,jaksch1998cold}, thus requiring the total simulation time $t_{\text{sim}}$ to fulfill $t_{\text{sim}} << t_{\text{coh}}$. Secondly, one needs to ensure that the atoms remain in the lowest Bloch band throughout the whole implementation. Hence, it is of crucial importance to shape the lattice and move the atoms in an adiabatic way. This is particularly important in our simulation scheme, where the auxiliary atoms have to be moved around or when the matter lattice has to be deformed to allow tunneling. This means that the corresponding timescale $t_{\text{mov}}$ should be bigger than the inverse of the frequency $\omega$ associated to the energy difference between lowest and first excited Bloch band ($t_{\text{mov}} >> 1/\omega$), while at the same time the obvious constraint $t_{\text{mov}} << t_{\text{coh}}$ has to be fulfilled. However, such techniques have also become well-controlled \cite{aguado2008creation,jaksch1998cold}. \\
Errors more specific to this proposal are connected with the tailoring of the two-body scattering. This requires a high degree of control over the overlap of the atomic wave functions and accurate timing of interaction during these collisions. This is also dependent on the ability to design and manipulate the magnetic field gradient in a precise manner.

\section{Summary}
In this work, two main results were discussed. First, a digital simulation scheme was proposed to realize lattice gauge theories in 3+1 dimensions including dynamical fermions using only two-body interactions. Its main feature is the ability of obtaining the magnetic plaquette interactions without using fourth-order perturbation theory, thus resulting in stronger interactions and allowing the study of wider phase-space regions compared to analogue approaches. Second, following the aforementioned simulation scheme, an implementation of a lattice gauge theory with a non-abelian gauge group - the dihedral group $D_{3}$ - was proposed, using ultracold atoms in optical lattices. Since the time evolution is performed in a Trotterized manner, intrinsic errors occur. These were studied in detail as a good bound on the trotter error gives more leeway to experimental errors. \\ 
The key ingredient of the digital simulation scheme is an auxiliary system which can be entangled with the physical system. This allows to create an isometry which mediates the complicated three and four-body terms of lattice gauge theory via the auxiliary system by using two-body interactions, as desired for implementations with various quantum simulation platforms. Moreover, it should be emphasized that all time evolutions in this algorithm are individually gauge invariant. The corresponding gauge group has to be either a compact Lie group or a finite group which is not a restriction for all relevant theories. In the case of compact Lie groups, the local Hilbert spaces of the gauge fields have an infinite dimension and therefore need to be truncated for a feasible implementation. However, since the isometry is defined in terms of the group element basis, the truncation has to be done there as well and can not be done in the typically used representation basis (see Sec. \ref{chapter2}). Examples for such truncations are $\mathbb{Z}_{N}$ for $U(1)$ or - as proposed in this work - $D_{N}$ for $O(2)$. \\
For the implementation of the lattice gauge theory with dihedral group $D_{3}$ - isomorphic to the symmetric group $S_{3}$ - we exploited the group structure of $D_{3}$ as a semidirect product. This allowed us to represent the gauge fields by a tensor product of a three-level and a two-level system and thus simplified the implementation. The potential gain from this procedure would be even higher for more complicated gauge groups exhibiting a semidirect product structure. \\
No sophisticated experimental techniques (e.g. Feshbach resonances) are required. However, precise control over atomic collisions is needed in order to obtain the desired time evolution, in particular gates entangling the auxiliary system with the physical system, as they do not depend on the simulated time and are thus more prone to experimental errors. \\
Future efforts on experimental techniques can therefore be targeted at the controllability of the relevant parameters, i.e. in particular fine tuning of the overlap integrals and the interaction time during scattering processes. The generation and experimental control of superlattices is important as well in order to create a staggering potential for the dynamical fermions. Also conducting experiments on simpler models - as currently set up for the Schwinger model - is a promising direction as it can serve as a proof of principle for the validity of quantum simulations of lattice gauge theories and might encourage more work in this direction. \\
From the theoretical point of view, a logical next step is to think of possibilities to realize more complicated gauge groups. One step towards that goal is to find suitable ways to truncate compact gauge groups like for example $SU(2)$ in a meaningful manner. 

\ack
J.I.C. is supported by the ERC QENOCOBA under the EU Horizon 2020 program (grant agreement 742102).

\appendix

\section{Details on $D_{N}$ lattice gauge theory}
In the following we will present some details on the lattice gauge theory of $D_{N}$. Due to its non-abelian gauge group the representations of the group become non-trivial and thus a lot of terms more complicated. Therefore, we will start by discussing the most important group properties and the irreducible representations of $D_{N}$. \\
$D_{N}$ is the symmetry group of rotations by $\frac{2 \pi}{N}$ in a two-dimensional plane and reflections along a certain axis (any axis passing through the center of rotations is possible). It can be characterized by the set 

\begin{equation}
D_{N} = \{ g= (p , m) \equiv R\left(2\pi/N \right)^p S^m | p \in [0,N-1 ) \hspace{2pt} \mathrm{and} \hspace{2pt} m \in \{ 0,1 \}  \} 
\end{equation}
The structure of the group is defined by the composition rules: 

\begin{equation}
(p,m) \cdot (r,n) = (p+(-1)^m r, m+n)  
\end{equation}
where the addition of p and r is understood as modulo $N$, respectively modulo 2 for $m$ and $n$. The neutral element is $e= (0,0)$ and the inverse element of $(p,m)$ is $(p,m)^{-1}=(p(-1)^{m+1}, m)$. The representation theory of $D_{N}$ ($N$ odd and $N \geq 3$) is characterized by the three irreducible representations shown in the table below: \\

\begin{center}
\begin{tabular} {|c|c|}
\hline
 
Trivial (dimension 1) & $D^{t}(p,m) = 1$ \\ \hline
Sign (dimension 1) & $D^{s} (p,m)=(-1)^m$ \\ \hline 
k-th (dimension 2) & $D^{k} (p,m)=e^{i\frac{2 \pi p}{N}k\sigma_{z}} \sigma_{x}^m  $ \\
\hline
\end{tabular} 
\end{center} 
We exclude the cases where $N$ is even, since they have additional sign representations and are not relevant for the discussion of $D_{3}$. With the above table, the electric Hamiltonian can easily be given in the representation basis. However, since this form of the Hamiltonian is not very feasible for the proposed quantum simulation we will show how to transform it to the group element states: 

\begin{equation} \label{Trace}
\begin{aligned}
h_{E}(\mathbf{x},k)&= \sum_{g,g'} \sum_{j,m,n} f(j) \ket{g} \braket{g | jmn} \braket{jmn | g'} \bra{g'} \\
&= \sum_{g,g'} \sum_{j} \frac{dim(j)}{|G|} f(j) \ket{g} \mathrm{Tr} (D^{j} (g) D^{j \dagger} (g')) \bra{g'}
\end{aligned}
\end{equation}
To specify this expression for $D_{N}$ we need to calculate the trace from above for all irreducible representations: \\ 
Trivial representation: $\mathrm{Tr} (D^{t} (p,m) D^{t \dagger} (p',m'))= 1$ \\
Sign representation: $\mathrm{Tr} (D^{s} (p,m) D^{s \dagger} (p',m')) = (-1)^{m+m'}$ \\
k-th representation: $\mathrm{Tr} (D^{k} (p,m) D^{k \dagger} (p',m'))= \delta_{mm'} (e^{i\frac{2 \pi}{N}k (p-p')} + e^{-i\frac{2 \pi}{N}k (p-p')})$ \\ \\
Inserting this into (\ref{Trace}) we obtain: 

\begin{equation} \label{Trace2}
\begin{aligned}
h_{E}(\mathbf{x},k)=& \frac{1}{2N} \sum_{p,p'} \sum_{m,m'} \ket{p,m} \bra{p',m'} \\
&\left( f_{t} + f_{s} (-1)^{m+m'} + 2 \sum_{k=1}^{N-1} f_{k}  \delta_{mm'} (e^{i\frac{2\pi}{N} k(p-p')} + e^{-i\frac{2\pi}{N}k (p-p')}) \right) \\ 
\end{aligned}
\end{equation}
The expression simplifies if we go to the conjugate basis of $\{ \ket{p} \}$ which can be viewed as the angular momentum basis $\{ \ket{l} \}$ characterized by the relation

\begin{equation}
\braket{l,m|p,n}=\frac{1}{\sqrt[]{N}} \delta_{mn} e^{-i\frac{2\pi}{N}lp}.  
\end{equation}
We obtain 
\begin{equation}
\begin{aligned}
h_{E}(\mathbf{x},k)&=\frac{1}{2 N^2} \sum_{m,m'} \sum_{l,l'} \ket{l,m} \sum_{p,p'} \left( f_{t}e^{-i\frac{2\pi}{N}lp} e^{i\frac{2\pi}{N}l'p'} + f_{s} e^{-i\frac{2\pi}{N}lp} e^{i\frac{2\pi}{N}l'p'} (-1)^{m+m'} \right. \\ 
&\left.+ 2 \sum_{k=1}^{N-1} f_{k}  \delta_{mm'} (e^{i\frac{2\pi}{N}(k-l)p} e^{i\frac{2\pi}{N}(l'-k)p'} + e^{-i\frac{2\pi}{N}(k+l)p} e^{i\frac{2\pi}{N}(k+l')p'}) \right) \bra{l',m'} \\ 
&=\frac{1}{2} \sum_{m,m'} \left(  f_{t} \ket{0,m} \bra{0,m'} + f_{s} (-1)^{m+m'} \ket{0,m} \bra{0,m'} +  2 \sum_{l \neq 0} f_{l} \ket{l,m} \bra{l,m} \right)  
\end{aligned}
\end{equation}
where the coefficients $f_{l}$ have to satisfy the constraint $f_{l}=f_{-l} \hspace{2pt} \forall l
$. If we redefine the coefficient for the trivial and sign representation as $f_{r} \equiv f_{t}-f_{s}$ and $f_{0} \equiv f_{s}$ we can simplify the expression further: 

\begin{equation}
\begin{aligned}
h_{E}(\mathbf{x},k)=\frac{1}{2} \sum_{m,m'}  f_{r} \ket{0,m} \bra{0,m'} + \sum_{l= -(N-1)/2}^{(N-1)/2} \sum_{m}  f_{l} \ket{l,m} \bra{l,m} \\
\end{aligned}
\end{equation}
The second term can be viewed as the electric energy of $Z_{N}$ as it acts trivially on the gauge field Hilbert space corresponding to reflections.

\section{Trotter errors}  \label{App:error}

For the bounds on the trotter error of the digital quantum simulation (presented in Sec. \ref{algo}) a computation of commutators and nested commutators of the different parts of the Hamiltonian is required. Since the calculation of these commutators for a general lattice gauge theory is very lengthy, it is only sketched here. 

\subsubsection{First order}
For the first order formula the ordinary commutators need to be computed. Starting with the commutator between gauge-matter interactions on different links $i$ and $j$, we obtain: 

\begin{equation}
\begin{aligned}
&\left[ H_{GM,i},H_{GM,j} \right] \\
=& \left[ \sum_{x, k_{i}} \lambda_{GM} \hspace{2pt} \psi_{m}^{\dagger}(x) U_{mn}(x,k_{i}) \psi_{n}(x+k_{i}) +h.c , \sum_{y, k_{j}} \lambda_{GM} \hspace{2pt} \psi_{m}^{\dagger}(y) U_{mn}(y,k_{j}) \psi_{n}(y+k_{j}) + H.c\right] \\
=&\lambda_{GM}^2 \sum_{x/\{ boundary \} } \psi_{n}^{\dagger}(x+k_{i})  U_{nm}^{\dagger}(x,k_{i}) U_{mn'}(x,k_{j}) \psi_{n'}(x+k_{j}) - H.c. \\
+&\psi_{m}^{\dagger}(x) U_{mn}(x,k_{i}) U_{nm'}^{\dagger}(x+k_{i}-k_{j},k_{j}) \psi_{m'}(x+k_{i}-k_{j}) - H.c. \\
=&\lambda_{GM}^2 \sum_{x/\{ boundary \} } \mathcal{U}_{W1}^{\dagger} \mathcal{U}_{W2}  (\psi_{n}^{\dagger}(x+k_{i})\psi_{n}(x+k_{j})- H.c.) \mathcal{U}_{W2}^{\dagger} \mathcal{U}_{W1} \\
+& \mathcal{U}_{W3} \mathcal{U}_{W4}^{\dagger} (\psi_{n}^{\dagger}(x)\psi_{n}(x+k_{i}-k_{j})- H.c.) \mathcal{U}_{W4} \mathcal{U}_{W3}^{\dagger}
\end{aligned}
\end{equation}
where we used the unitary operators $\mathcal{U}_{W}$ from Sec. \ref{algogauge} to reduce the gauge-matter terms to pure fermionic tunneling terms, thus allowing to estimate this expression: 

\begin{equation}
\begin{aligned}
\| \left[ H_{GM,i},H_{GM,j} \right] \| \le \lambda_{GM}^2 \frac{\mathcal{N}_{\mathrm{links}}}{d} d_{U}
\end{aligned}
\end{equation}
where $d_{U}$ is the dimension of the representation of $U$ under the gauge group and therefore the operator norm of the tunneling term. In the next step, the commutator between the matter- and gauge-matter interactions is calculated: 

\begin{equation}
\begin{aligned}
&\left[ H_{M},H_{GM,i} \right] \\
=&\left[ \sum_{x} M (-1)^{x} \psi^{\dagger}_{n'}(x) \psi_{n'}(x), \sum_{y,k_{i}}\lambda_{GM} \psi_{m}^{\dagger}(y) U_{mn}(y,k_{i}) \psi_{n}(y+k_{i}) + H.c \right] \\
=&2 \sum_{x} M \lambda_{GM} (-1)^{x} \psi_{m}^{\dagger}(x) U_{mn}(x,k_{i}) \psi_{n}(x+k_{i}) - H.c.
\end{aligned}
\end{equation}
We rewrite this expression again in terms of the unitary operators $\mathcal{U}_{W}$ which allows us to bound the commutator in the following way: 

\begin{equation}
\begin{aligned}
\| \left[ H_{M},H_{GM,i} \right] \| \le 2 M \lambda_{GM} \mathcal{N}_{\mathrm{links}} d_{U} 
\end{aligned}
\end{equation}
For the commutator with the electric part the whole gauge-matter Hamiltonian is considered as every part does not commute with $H_{E}$. To bound this expression from above we write $H_{GM}$ again in terms of the unitary operators $\mathcal{U}_{W}$, similar to the previous calculations and obtain: 

\begin{equation}
\begin{aligned}
&\| \left[ H_{GM},H_{E} \right] \| \leq \mathcal{N}_{\mathrm{links}} \lambda_{GM}  \lambda_{E} \max_{j} |f(j)| 2 d_{U} 
\end{aligned}
\end{equation}
The last commutator is the one between the magnetic and electric Hamiltonian. Since every link is contained in $2(d-1)$ plaquettes, the commutator is straightforwardly estimated as: 

\begin{equation}
\begin{aligned}
\| \left[ H_{B},H_{E} \right] \| \le \lambda_{B} \lambda_{E} \mathcal{N}_{\mathrm{links}} 8(d-1) \max_{j} |f(j)| d_{U} 
\end{aligned}
\end{equation}
 
\subsubsection{Second order}
For a bound on the second-order formula we need to calculate all nested commutators. The computations of them are done in the same manner as for the ordinary commutators, there are no additional tricks required. Since these calculations are very lengthy, we will just give the bounds obtained for each nested commutator: 

\begin{equation}
\begin{aligned}
&\| \left[[H_{B},H_{E}],H_{E}\right] \| \leq \lambda_{E}^{2} \lambda_{B} \max_{j} |f(j)|^2 \mathcal{N}_{\mathrm{links}} 64(d-1) d_{U} \\
&\| \left[[H_{B},H_{E}],H_{B}\right] \| \leq \lambda_{E} \lambda_{B}^2 \max_{j} |f(j)|  \mathcal{N}_{l\mathrm{links}}  64(d-1)^2 d_{U}^2 \\
&\| \left[\left[H_{E},H_{GM}\right],H_{GM} \right] \| \leq \lambda_{GM}^2 \lambda_{E} \max_{j} |f(j)| \mathcal{N}_{\mathrm{links}} (2(2d-1)+1) 4d_{U}^2 \\
&\| \left[\left[H_{E},H_{GM}\right],H_{E} \right] \| \leq \lambda_{GM} \lambda_{E}^2 \max_{j} |f(j)|^2  \mathcal{N}_{\mathrm{links}} 4d_{U} \\
&\| \left[\left[H_{M},H_{GM}\right],H_{GM} \right] \| \leq \lambda_{GM}^2 M \mathcal{N}_{\mathrm{links}} 8d d_{U} \\
&\| \left[\left[H_{M},H_{GM}\right],H_{M} \right] \| \leq 4\lambda_{GM}M^2 \mathcal{N}_{\mathrm{links}} d_{U} 
\end{aligned}
\end{equation}
In the last step the nested commutator among the different gauge-matter Hamiltonians needs to be computed: 

\begin{equation}
\begin{aligned}
&\| \left[\left[H_{GM,i},H_{GM,j}\right],H_{GM,l} \right] \| &\leq \lambda_{GM}^3 \frac{\mathcal{N}_{\mathrm{links}}}{d} 2d_{U} 
\end{aligned}
\end{equation}
To obtain the error bound for the whole gauge matter interactions we need to calculate how many times the commutator from above appears. There are $2d$ different gauge-matter Hamiltonians which are implemented separately. Recalling the second order formula, this gives rise to two partial sums over the natural numbers: 

\begin{equation}
\begin{aligned}
& \sum_{k=1}^{2d-1} \| [[H_{GM,k},H_{GM,k+1}+..+H_{GM,2d}],H_{GM,k+1}+..+H_{GM,2d}] \| \\
&+ \frac{1}{2} \| [[H_{GM,k},H_{GM,k+1}+..+H_{GM,2d}],H_{GM,k}] \| \\ 
&\le \lambda_{GM}^3 \frac{\mathcal{N}_{\mathrm{links}}}{d} 2d_{U} \sum_{x=1}^{2d-1} x^2 + \frac{x}{2} =\lambda_{GM}^3d_{U} \mathcal{N}_{\mathrm{links}}(2d-1) \left(\frac{2}{3} (4d-1)+1 \right) 
\end{aligned}
\end{equation}
Inserting all these commutators into the formulas of the total trotter error will then result in the bounds given in Sec. \ref{algo}.

\section*{References}
\bibliographystyle{iopart-num.bst}
\bibliography{bibliography}

\end{document}